\documentclass[11pt]{article}
\usepackage{geometry}
\usepackage{cite}
\usepackage{amssymb,amsmath}
\usepackage{graphicx}

\renewcommand{\vec}[1]{\mathbf{#1}}
\newcommand{\sunobj}[1]{\pmb{#1}}
\newcommand{\atmidp}[1]{\overline{#1}}
\newcommand{\nnb}[1]{\langle #1 \rangle}
\DeclareMathOperator{\trace}{Tr}
\DeclareMathOperator{\diag}{diag}
\DeclareMathOperator{\ce}{ce}
\DeclareMathOperator{\se}{se}
\newcommand{\eulere}{\mathrm{e}}

\begin{document}
\newsavebox{\prepno}
\savebox{\prepno}{\parbox{5cm}{UMIST/Phys/TP/00-1\\ECT*-00-001}}
\markboth{\usebox{\prepno}}{\usebox{\prepno}}

\title{Towards a Many-Body Treatment of Hamiltonian Lattice ${SU}(N)$
Gauge Theory}

\author{N.E. Ligterink\thanks{Present address: ECT*, Strada delle Tabarelle 286,
I-38050 Villazzano (Trento), Italy.},
 N.R. Walet
 and R.F. Bishop\\
 {\it Department of Physics, UMIST, PO Box 88, Manchester, M60 1QD, United Kingdom}\\
E-mail: ligterin@ect.it,
N.R.Walet@umist.ac.uk,
R.F.Bishop@umist.ac.uk}

\maketitle
\thispagestyle{myheadings}

\begin{abstract}
We develop a consistent approach to Hamiltonian lattice gauge theory,
using the maximal-tree gauge. The various constraints are discussed and
implemented. An independent and complete set of variables for the
colourless sector is determined. A general scheme to construct the
eigenstates of the electric energy operator using a symbolic method is
described.  It is shown how the one-plaquette problem can be mapped
onto a $N$-fermion problem. Explicit solutions for $U(1$), $SU(2)$,
$SU(3)$, $SU(4)$, and $SU(5)$ lattice gauge theory are shown.
\end{abstract}

\maketitle


\noindent
{\Large \emph{Contents}} \\
\vspace{3mm}
\noindent
\begin{trivlist}
\item[1.] \emph{Introduction.}
\item[2.] \emph{Hamiltonian lattice gauge theory.} {2.1}.~Discretisation.
{2.2}. Gauss' law.
\item[{3}.] \emph{Explicit Hamiltonian.}
{3.1}.~Conditions on variables.
{3.2}.~The gauge-fixed Hamiltonian.
{3.3}.~Boundaries.
\item[{4}.] \emph{Further constraints.}
\item[{5}.] \emph{The one-plaquette problem.}
{5.1}.~Angular representation.
{5.2}.~$\trace [U]$ as a raising operator.
{5.3}.~The harmonic approximation.
{5.4}.~Large-$N$ limit.
{5.5}.~Explicit solutions.
\item[{6}.]\emph{Wave functionals.}
\item[{7}.] \emph{Conclusions.}
\item[Appendix A.] \emph{One-dimensional many-fermion systems and their representations}
{A.1} The magnetic term.
\item[Appendix B.] \emph{Equivalent forms of the electric energy.}
\end{trivlist}
\newpage

\section{Introduction}

Strongly interacting gauge-field theories, where perturbation theory
is of no help, have been a long-standing problem. The theory of strong
interactions, quantum-chromodynamics (QCD), is probably the most
well-known theory of this type.
This paper is concerned with one approach to such problems, 
the Hamiltonian description of the lattice
version of  gauge-field theory (LGFT). For most investigations in this area this
is not the method of choice; the majority of calculations on such
theories are done within the Lagrangian formalism. The main reason for
this is the fact that, at least in principle, the Lagrangian method is
elegant and simple. It is based on a path-integral approach
to the imaginary-time propagation. Much work has been done to improve
the accuracy of the Monte Carlo method used in the numerical
implementation of the Lagrangian approach, further increasing the
viability of the method.

Thus, anyone wishing to pursue a different path needs first to build a
strong case for the merits of their approach. For the form of the
Hamiltonian approach we shall consider in this paper we see four
important advantages:

\begin{enumerate}

\item The Lagrangian approach, being based on an imaginary time
evolution, does not allow easy access to the vacuum wave functional.
By contrast, in the Hamiltonian approach such a wave functional is at
the core of the calculation, and we cannot avoid to calculate it.
Once the vacuum wave functional is known, most properties of theories
such as QCD, like confinement and chiral symmetry breaking, should
follow automatically.

\item Time-dependent phenomena can only be discussed in a real time  
(Hamiltonian) setting.

\item  The physical interpretation of the variables is more
transparent. For example, electric and magnetic operators have their
classical meaning. 

\item The Hamiltonian approach can studied not only with numerical, 
but also with analytical techniques, so we can study low-lying excitations
with a harmonic approximation and we can disentangle the dependence of
observables on the parameters in the approximation.

\end{enumerate}

These advantages are accompanied by important disadvantages.  Due to
the fact that the Hamiltonian requires a quantisation plane to be
defined, we lose the manifest Lorentz symmetry, that arises so
naturally in the Lagrangian approach. In the full solution to the
Hamiltonian problem this should be recovered, but approximations can
introduce spuriosities. Furthermore we have to fix the gauge, and the
degrees of freedom in the Hamiltonian formulation of the theory will
then naturally depend on the choice of gauge. The issue of gauge fixing is
a rather non-trivial exercise, as we shall see in this paper.

We shall first, in Sec.~\ref{sec:Hlatg}, give a short introduction to
our version of lattice gauge theory. In Sec.~\ref{sec:Ham} we then
discuss how we can use the idea of a maximal tree to define a fully
gauge-fixed Hamiltonian, which is derived in an explicit form.  In
Sec.~\ref{sec:FC} we discuss those constraints relating the trace of
powers of $SU(N)$ matrices, and we define a symbolic method to
construct eigenvalues of the electric energy operator, which should be
useful in any approach to Hamiltonian LGFT.  We then study the one
plaquette problem in Sec.~\ref{sec:OP}, first in its generality,
showing how we can derive a simple $N$-fermion problem equivalent to
the $SU(N)$ one-plaquette problem. We discuss both the weak-coupling
(harmonic) and the large-$N$ limits. Explicit results for the one
plaquette spectra as function of the coupling constant are shown for
$U(1)$, and $SU(N)$ LGFT's, with $2 \leq N \leq 5$.  Finally we
discuss the results and future directions in Sec.~\ref{sec:concl}.

\section{Hamiltonian lattice gauge theory\label{sec:Hlatg}}

The degrees of freedom in a gauge theory, such as electromagnetism,
can be divided into two sets. The first set consists of the charges
and currents in the system together with the associated static
electric and magnetic fields. The second set contains the
electromagnetic waves. However, since a charge in a moving frame is a
current, relativistic coordinate transformations mix electric and
magnetic variables.  Moreover, since the velocity of electromagnetic
waves is constant, the two polarisation directions, perpendicular to
the propagation direction, also undergo complicated transformations
under a change of frame. Therefore a relativistically invariant
formulation of electromagnetism is very useful.

As is well known, the Lagrangian of electromagnetism, without static sources,
\begin{equation}
\mathcal{L}=\frac{1}{2}\sum_{i=1}^3(E_{i}^{2}-B_{i}^{2})\quad ,
\end{equation}
can be formulated in terms of a covariant four-vector potential $A_{\mu }$
as 
\begin{equation}
\mathcal{L}=\frac{1}{4}(\partial _{\mu }A_{\nu }-\partial _{\mu }A_{\nu })^{2}\quad ,
\end{equation}
where $B_{i}=\epsilon _{ijk}\partial _{j}A_{k}$ and $ E_{i}=\partial
_{i}A_{0}-\partial _{0}A_{i}$.  We use a notation in which roman
indices $i,j,k,\ldots$ run over the spatial, while greek indices
$\mu,\nu,\ldots$ run over the values $0,1,2,\ldots,D$, where $D$ is the 
number of space dimensions and the temporal axis has index 0. 
The introduction of the vector potential, or gauge
field, serves a number of purposes. Firstly, since the gauge field
transforms as a vector, the Lagrangian is now manifestly
relativistically invariant. Secondly, it is formulated in terms of
canonical variables $A_{\mu }$ and $\partial _{0}A_{\mu }$ which
should make quantisation more straightforward; and thirdly, the two
constraints for the electric and the magnetic fields, the source-free
Maxwell equations $\partial_{i}B_{i}=0$ and $\partial_{0}B_{i} =
-\epsilon _{ijk}\partial _{j}E_{k}$, are automatically satisfied. The
price we pay for this is the appearance of superfluous degrees of
freedom, the gauge, and an entanglement of the sources giving rise to
these variables.

The original Abelian gauge theory of electromagnetism was extended by
Yang and Mills \cite{YangMills} using gauge fields of more complicated
structure, including internal degrees of freedom. This generates
self-interactions since the gauge fields do not commute, but are
chosen to obey the commutation relations of a specific Lie algebra. We
shall concentrate on a gauge-field Lagrangian where the field $\sunobj
{A}_{\mu }$ is an element of the Lie algebra ${su}(N)$,
\begin{equation}
\mathcal{L}=\frac{1}{2g^{2}}\trace [\sunobj {F}_{\mu \nu }\sunobj
{F}^{\mu \nu }]\quad .
\end{equation}
Here the field tensor is the skew-Hermitian matrix $\sunobj {F}_{\mu
\nu }\equiv \partial _{\mu }\sunobj {A}_{\nu }-\partial _{\nu }\sunobj
{A}_{\mu }-i[\sunobj {A}_{\mu },\sunobj {A}_{\nu }]$, and the field
variable $\sunobj {A}_\mu$ is an element of the algebra, conveniently
parametrised as 
\begin{equation}
\sunobj {A}_{\mu }\equiv g\frac{1}{2}\lambda^{a}A^{a}_{\mu }\quad.
\label{eq:paramA}
\end{equation}
The $\lambda^a$ are the $N^2-1$ generators of the Lie algebra,
satisfying the commutation relations
\begin{equation}
[\lambda ^{a},\lambda ^{b}]=
{2}i f^{abc}\lambda ^{c}\quad.
\end{equation}
The index $a$ thus runs from $1$ to $N^2-1$. The $\lambda^a$ can be
represented by traceless $N\times N$ matrices, normalised such that
their squares have trace 2, as can be seen from the anticommutation
relations
\begin{equation}
\{\lambda ^{a},\lambda ^{b}\}=
{2}d^{abc}\lambda ^{c}+
\frac{4}{N}\delta_{ab} I\quad.
\end{equation}
In Eq.~(\ref{eq:paramA}) we have absorbed the coupling constant $g$ in
the field $\sunobj{A}_{\mu}$, so that we can interpret the fields
geometrically, since the field tensor is now the curvature that
follows from the covariant derivative
\begin{equation}
\sunobj{d}_{\mu }\equiv \partial _{\mu }-i[\sunobj{A}_{\mu }, \cdot ] \quad,
\end{equation}
i.e.,
\begin{equation}
 [\sunobj{d}_\mu,\sunobj{d}_\nu] =
[ \sunobj{F}_{\mu\nu}, \cdot ] \quad .
\end{equation}

As we are interested in the Hamiltonian, we perform the standard
equal-time quantisation and reformulate the Lagrangian in generalised
electric and magnetic fields. This is strictly speaking a
3-dimensional result, since this interpretation requires the use of
three-dimensional algebra. We shall nonetheless use the result below for other
dimensions as well. We find
\begin{equation}
\mathcal{L}=\frac{1}{g^{2}}\trace \left[ \sum _{k=1}^{D}(\sunobj E_{k}^{2}-
\sunobj B_{k}^{2})
\right] \quad,
\end{equation}
where $\sunobj B_{i}=-\frac{1}{2}\epsilon _{ijk}\sunobj F_{jk}$ and
$\sunobj E_{i}= \sunobj F_{i0}$. Since we wish to impose the temporal gauge
$\sunobj A_0 = 0$, we separate the Hamiltonian in two parts, thereby isolating
the $\sunobj A_0$ dependent part: \cite{FS80}
\begin{equation}
\mathcal{L}=\frac{1}{g^{2}}\trace \left[ \sum _{k=1}^{D}(\sunobj E_{k}^{2}-
\sunobj B_{k}^{2}) \right]_{\sunobj A_0 = 0} + \frac{1}{g^{2}}\trace 
\left[ \sunobj A_{0}\sunobj {\mathcal{G}}+
\sunobj A_{0}\sunobj {\mathcal{X}}(\sunobj A_{0})\right] \quad,
\end{equation}
where we added a total divergence.
The function $\sunobj {\mathcal{X}}$ is second order in $\sunobj
A_{0}$ and does not contribute to the equations of motion, or the
constraint equations in the temporal gauge (${\sunobj A}_{0}=0$)
discussed below. Since the time-derivative of $\sunobj A_{0}$ does not
occur in the Lagrangian, the (Lagrangian) extremal action variational
principle leads to an equation of motion for $\sunobj A_{0}$ that is a
time-independent algebraic equation, which shows that the $\sunobj
A_0$ takes on a time-independent constant value. This set of equations
(one for each colour index) are the non-Abelian analogue of the Gauss'
law constraint, which in the absence of colour charges take the simple
form
\begin{equation}
{\mathcal{G}}^{a}(\vec x)=0\quad,
\label{gss1}
\end{equation} 
where
\begin{equation}
{\mathcal{G}}^{a}(\vec{x})=\sum _{i=1}^{D}[\partial _{i}
E^{a}_{i}(\vec{x})+g f^{abc}A_{i}^{b}(\vec{x})
E_{i}^{c}(\vec{x})] = \sum_{i=1}^D \sunobj d_i E_i^a\quad.
\label{gss2}
\end{equation}
The components of the fields can be obtained via the 
relation 
\begin{equation}
A^{a}_{\mu }=\frac{1}{g}\trace [\sunobj A_{\mu }\lambda^{a}], \quad
E^{a}_{\mu }=\frac{1}{g}\trace [\sunobj E_{\mu }\lambda^{a}]\quad .
\end{equation}
The constraints obey the same commutation relations as the generators
of the gauge group. Thus Gauss' law cannot be implemented as a strict
operator condition as it leads to contradictions, since the
non-commuting constraints cannot all be diagonalised simultaneously.
However, within the physical (in this case colourless) subspace
defined by
\begin{equation}
\mathcal{G}^{a}(\vec{x})|\mathrm{Phys}\rangle =0\quad ,
\end{equation}
no such problem arises, since the eigenvalue of the commutators is
also $0$. The space of states consists of wave functionals, which are
functionals taking values on the $SU(N)$ group manifold.
We find $N^{2}-1$ functional conditions on each wave functional,
consisting of functions on the group manifold at each space point.

As is well known, quantisation of problems involving redundant degrees
of freedom (i.e., where some of the equations of motion are
constraints) is quite involved. The two main techniques used are Dirac
and BRS quantisation, and they require a large amount of additional
analysis.  For more details one can consult the seminal work by Dirac
\cite{Dir67,Dirac}, as well as Refs.~\cite{FS80,Mut87,HT92,KS75,Kog79}.

If we are able to work within the  physical subspace
only, one can ignore these formal problems and define the quantisation
of the canonical momenta
$\Pi _{i}^{a}=\partial _{0}A_{a}^{i}$ by
\begin{equation}
\Pi ^{a}_{i}(\vec{x})=\hat{E}_{i}^{a}=-i\frac{\delta }{\delta A_{i}^{a}(\vec{x})}\quad,
\end{equation}
which involves a functional derivative \cite{Ber66,Fel81} with respect to the
field variables.

Since $A^{a}_{0}$ is not dynamical, we cannot associate a canonical momentum
with it. We therefore use the temporal gauge, $A^{a}_{0}=0$, which leaves
us with residual gauge freedom $\sunobj {\phi }(\vec{x})$ independent of
the time coordinate, i.e., under the transformation 
\begin{eqnarray}
^{\sunobj {\phi }}\sunobj A_{\mu }(\vec{x}) & = & 
\sunobj {\phi }(\vec{x})\sunobj A_{\mu }(\vec{x})\sunobj {\phi }^{-1}(\vec{x})+
i[\partial _{\mu }\sunobj {\phi }(\vec{x})]\sunobj {\phi }^{-1}(\vec{x})\quad,\\
^{\sunobj {\phi }}\sunobj F_{\mu \nu }(\vec{x}) & = & 
\sunobj {\phi }(\vec{x})\sunobj F_{\mu \nu }(\vec{x})\sunobj {\phi }^{-1}(\vec{x})\quad,
\end{eqnarray}
where $\sunobj {\phi }\in SU(N)$, the Lagrangian is invariant.

\subsection{Discretisation}

Field theories suffer from singularities, both in the infrared and
ultraviolet limits. In many interesting cases, such as QCD
\cite{'tHooft}, these are renormalisable. Rather than dealing directly
with the continuum, we shall regularise the problem by introducing a
simple hypercubic lattice in the $D$-dimensional space, with lattice
spacing $a$. Since we are pursuing a Hamiltonian approach, time will
remain continuous. In this paper we shall concentrate on pure gauge
theory, without explicit charges (i.e., quarks).  In this case the
systems is described by a set of gauge fields $A^{a}_{i}$ at each
point of the $D$-dimensional lattice.  It looks like these could in
principle carry $D(N^{2}-1)$ degrees of freedom, as the group
${SU}(N)$ has $N^{2}-1$ generators $\lambda ^{a}$. However, even after
restricting the gauge fields by imposing the temporal gauge
$\sunobj{A}_0=0$, the Lagrangian is still invariant under a limited
set of gauge transformations that do not violate the temporal
constraint. Thus the apparent number of degrees of freedom is still
larger than that required to specify the dynamics uniquely. These
superfluous degrees of freedom arise since we have not fully exploited
the Gauss' law constraints.

In order to analyse this problem fully we first investigate the structure
of the Hamiltonian on the lattice. The gauge  fields 
\begin{equation}
\sunobj {A}_{i}=g\frac{1}{2}\sum _{a=1}^{N^{2}-1}\lambda ^{a}A^{a}_{i}\quad,
\end{equation}
are Hermitian, since $\lambda ^{a}$ is Hermitian.  In the
Schr\"odinger (wave function) representation the effect of these
fields can be incorporated as a non-Abelian change of phase of the wave
function between different points, a simple generalisation of the
Aharonov-Bohm effect in QED. This geometric interpretation can be
represented by a group element, a unitary transformation summing all
the small phase changes along the path connecting the two points,
\begin{equation}
U(\vec{x}_1,\vec{x}_2)\equiv P\exp \left\{ i\int
_{\vec{x}_1}^{\vec{x}_2}\sunobj {A}_{\mu }(\vec x)dx^{\mu
}\right\} \quad, \label{eq:pathprod}
\end{equation}
where we have introduced a path-ordered product, denoted by $P$, since
the quantities $\sunobj {A}_{\mu }(\vec x)$ for different values of
$\vec x$ are non-commuting matrices.

In practice we will only use this unitary transformation (which for
obvious reasons is also called a parallel transporter) on a link
between two nearest-neighbour lattice points $\vec{i}$ and $\vec{j}$
where $\vec a_{\mu }=\vec{x}_{\vec{i}}-\vec{x}_{\vec{j}}$ is a primitive
lattice vector in the direction $\mu $ (we shall use the bracket
notation $\langle\rangle$ to denote nearest-neighbour pairs)
\begin{eqnarray}
U_{\nnb{\vec{i}\vec{j}} } & = & 
P\exp \left\{ i\int _{\vec{x}_{\vec{i}}}^{\vec{x}_{\vec{j}}}\sunobj {A}_{\mu }(\vec x)dx^{\mu }\right\} 
\equiv \exp \left\{ i a_{\mu }
\atmidp{\sunobj {A}}_{\mu }( \vec{x}_{\nnb{\vec{i}\vec{j}}})\right\} 
\label{eq:defAmid}
\quad,\\
U_{\nnb{\vec{j}\vec{i}}} & = & 
\exp \left\{ - i a_{\mu }\atmidp{\sunobj {A}}_{\mu }(\vec{x}_{\nnb{\vec{i}\vec{j}}})\right\} 
=U^{\dagger }_{\nnb{ \vec{i}\vec{j}}}\quad.
\end{eqnarray}
In Eq.~(\ref{eq:defAmid}) we have defined
$\atmidp{\sunobj{A}}_{\mu }$ at the midpoints of the link, 
\begin{equation}
\vec{x}_{\nnb{\vec{i}\vec{j}}}=\frac{1}{2}(\vec{x}_{\vec{i}}+\vec{x}_{\vec{j}})\quad.
\end{equation}
Since $\sunobj {A}_{\mu }$ is an element of the ${su}(N)$ Lie algebra,
$U$ is a ${SU}(N)$ matrix, and satisfies $U^{-1}=U^{\dagger }$ and
$\det U=1$. The quantity
$\overline{\sunobj{A}}_{\mu}(\vec{x}_{\nnb{\vec{i}\vec{j}}})$ is a
geometrical average of $\sunobj {A}_{\mu }$ along the link $\nnb{
\vec{i}\vec{j}} $ in direction $\mu $. Since all future discussions
will only concern this averaged field, we suppress the bar from here
onwards.

\begin{figure}
\centerline{\resizebox*{3cm}{!}{\includegraphics{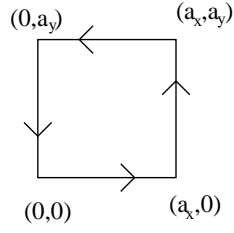}} }
\caption{The path for the elementary plaquette.}
\label{fig:plaquette}
\end{figure}

The four links around a primitive square (usually called plaquette) on
the lattice will give, when we take the trace, the simplest
gauge-invariant quantity on the lattice, and for that reason is often
used in lattice problems. As an example we take a plaquette in the
$xy$ plane, $(0,0)\rightarrow(a_{x},0)
\rightarrow(a_{x},a_{y}) \rightarrow(0,a_{y}) \rightarrow(0,0)$, see
Fig.~\ref{fig:plaquette}.  This leads to (the notation $\vec a_i$ is
used as a shorthand for $a_i$ times a unit vector in the direction
$i$)
\begin{eqnarray}
U_\square & = & 
e^{ia_{x}\sunobj A_{x}(\vec{a}_{x}/2)}
e^{ia_{y}\sunobj A_{y}(\vec{a}_{x}+\vec{a}_{y}/2)}
e^{-ia_{x}\sunobj A_{x}(\vec{a}_{y}+\vec{a}_{x}/2)}
e^{-ia_{y}\sunobj A_{y}(\vec{a}_{y}/2)}\nonumber\\
 & =  
& \exp\left\{i a_{x}a_{y}\left(\Delta _{y}\sunobj A_{x}(\vec{m})-
     \Delta _{x}\sunobj A_{y}(\vec{m})-
      [\sunobj A_{x}(\vec m),\sunobj A_{y}(\vec m)]\right)+
     {\mathcal{O}}(a^{3}\right\}\quad,
\end{eqnarray}
where $\vec m=(\vec a_{x}+\vec a_{y})/2$ is the centre of the plaquette,
and we have assumed that both $a_x$ and $a_y$ are of order $a$.
We have used the Campbell-Baker-Hausdorff formula to combine the exponentials, 
\begin{equation}
\exp(A)\exp(B)=\exp \left\{A+B+\frac{1}{2}[A,B]+\frac{1}{12}[A,[A,B]]+\frac{1}{12}[[A,B],B]+\cdots \right\}\quad.
\end{equation}
The operator $\Delta _{i}$ is the lattice derivative, defined as 
\begin{equation}
\Delta _{i}\sunobj A_{j}(\vec m)=
\frac{\sunobj A_{j}(\vec m+\vec a_{i}/2)-\sunobj A_{j}(\vec m-\vec a_{i}/2)}{a_i}\quad.
\end{equation}
%
Eventually we find, using $a_{x}=a_{y}=a$, 
\begin{equation}
U=e^{-a^{2}i \sunobj F_{xy}+{\mathcal{O}}(a^{3})}\quad.
\end{equation}
We would like to associate the trace of $U$ with a lattice version of
the QCD Lagrangian. The requirement is that in the $a\rightarrow 0$
limit the Lagrangian becomes the continuum one. Thus we must analyze
the expansion of the trace.  If we expand $U_\square$ in a Taylor
series in $a$, we find that the $a^2$ and $a^3$ terms are traceless,
and so are all the terms of the order $a^4$ except for the term $-i
(\sunobj F_{xy})^2$. Thus it seems natural to define $\trace U_\square$ as
the lattice potential. Unfortunately, this is a complex
quantity, and we need to combine it with its complex conjugate to get
a real result,
\begin{equation}
\frac{1}{2 a^4g^{2}} \trace [\sunobj F_{xy}\sunobj F_{xy}]=
\frac{1}{2 a^4g^{2}}\trace [2-U_\square-U_\square^{-1}]+{\mathcal{O}}(a)\quad.
\end{equation} 
Clearly this approaches the continuum result as $a\rightarrow 0$.
This is the one-plaquette contribution to the magnetic part of the
Wilson action \cite{WilsonA,MM94}.

In a Hamiltonian formalism we need to construct canonical momenta.
The natural choice is the differential realisation of the generators
at each lattice point (i.e., the generalised angular momentum
operators for the group ${SU}(N)$). These act naturally on the
original lattice variables $\sunobj{A}_{\mu} (\vec{x}_{\vec{i}})$, but
as stated above we really would like to formulate the problem in terms
of link variables $U_{\nnb{\vec{i}\vec{j}}}$.  Using the definition of
$U_{\nnb{\vec{i}\vec{j}}}$ as a path-ordered product,
Eq.~(\ref{eq:pathprod}), we find that the action of the momenta,
conventionally called electric fields $E$, is given by
\begin{eqnarray}
E_{\mu }^{a}(\vec x_{\vec{i}})U_{\nnb{\vec{i}\vec{j}}}&=&
-\frac{1}{2}\lambda ^{a}U_{\nnb{\vec{i}\vec{j}}}\quad,
\nonumber\\
E_{\mu }^{a}(\vec x_{\vec{i}})U_{\nnb{\vec{j}\vec{i}}}&=&
\frac{1}{2}U_{\nnb{\vec{j}\vec{i}}}\lambda ^{a}\quad,
\end{eqnarray}
where $\vec{j} = \vec{i + e}_\mu $, so $\mu$ must point along the link
 $\nnb{\vec i \vec j}$ in the positive direction. The second
relation follows from the Hermitian conjugation. This can
also be written as a differential operator acting on the link-averaged
gauge fields $A^{a}_{\mu }(\vec x_{\nnb{\vec{i}\vec{j}}})$,
\begin{equation}
E_{\mu }^{a}(x_{\vec{i}})=-\frac{i}{a_{\mu }}\frac{\delta }{\delta
A^{a}_{\mu }(\nnb{\vec{x}_{\vec{i}\vec{j}}})}
\delta_{\vec{x}_{\vec{j}}-\vec{x}_{\vec{i}},\vec{a}_{\mu }}\quad.
\end{equation}
We shall also use the notation $E^a_{\vec i \vec j}$ for this same operator.

Ignoring problems with overcompleteness, we can derive the
Kogut-Susskind Hamiltonian, \cite{KS75}
\begin{eqnarray}
H_{\mathrm{KS}} & = &
\frac{g_{t}^{2}}{2a^{D-2}} \left( \sum _{\vec{i},\mu }^{Dn^{D}}
	\sum _{a=1}^{N^{2}-1}E_{\mu }^{a}(\vec{x}_{\vec{i}})E^{a}_{\mu }(\vec{x}_{\vec{i}})+
\frac{1}{g^3_{t}g_{s}a^{6-2 D}}
\sum _{\nnb{ \vec{i}\vec{j}} }^{\frac{1}{2}D(D-1)n^{D}}
\trace [2-U(\vec{m}_{\vec{i},k,l})-U^{-1}(\vec{m}_{\vec{i},k,l})] \right) 
\nonumber  \\
& = & \frac{g_{t}^{2}}{2a^{D-2}} \left( H_E + \frac{1}{g^3_{t}g_{s}a^{6-2 D}} H_M \right) \quad,
\label{eq:HKS}
\end{eqnarray}
where $n$ is the length of the lattice, and $D$ is the number of space
dimensions.  The quantity $U(\vec{m}_{\vec{i},k,l})$ is the Wilson
plaquette around the face centre $\vec{m}_{\vec{i}}$ in the directions
$k$ and $l$,
\begin{equation}
U(\vec{m}_{\vec{i},k,l})=
U_{\vec{i},(\vec{i}+\vec a_{k})}
U_{(\vec{i}+\vec a_{k}),(\vec{i}+\vec a_{k}+\vec a_{l})}
U_{(\vec{i}+\vec a_{k}+\vec a_{l}),(\vec{i}+\vec a_{l})}
U_{(\vec{i}+\vec a_{l}),\vec{i}}\quad,
\end{equation}
where $\vec{m}_{\vec{i}}=\vec{x}_{\vec{i}}+\frac{1}{2}(\vec a_{k}+\vec
a_{l})$.  We have written down the slightly more general form of
Hamiltonian obtained when we realise that the coupling constants
$g_{t}$ and $g_{s}$ for the spatial and time components of $\sunobj{A}$
can be different, since the Hamiltonian breaks manifest Lorentz
invariance. In principle one has to solve a renormalisation problem to
find the physical couplings that correspond to these bare ones, a
problem that we shall not address in this paper.

\subsection{Gauss' law\label{sec:Gauss}}
The constraints, Eqs.~(\ref{gss1},\ref{gss2}), can also be interpreted
as arising from the invariance of the gauge-fixed Lagrangian under 
residual gauge transformations $\sunobj \phi$. Unfortunately,
the result (\ref{gss2}) is only correct for continuum theories,
since it involves the field $\sunobj A_\mu$ rather than its lattice analogue 
$U$. The derivation of Gauss' law constraints on the lattice is discussed in
Ref.~\cite{Baa85}. The approach is based on the fact that  a 
wave functional, which dependens on the overcomplete set of link variables,
\begin{equation}
\Psi( \{U \} ) \quad ,
\end{equation}
Should be invariant under a gauge transformation on
a single lattice site,
\begin{eqnarray}
U_{\vec i, \vec i + e_\mu} & \to & e^{i \sunobj \phi_{\vec i}} 
U_{\vec i, \vec i + e_\mu} e^{i \sunobj \phi^\dagger_{\vec i + e_\mu}} \\
U_{\vec i - e_\mu, \vec i} & \to & e^{i \sunobj \phi_{\vec i- e_\mu}} 
U_{\vec i - e_\mu, \vec i} e^{i \sunobj \phi^\dagger_{\vec i }} \quad .
\label{eq:gaussinvb}
\end{eqnarray}
The invariance can be written in terms of generators as
\begin{equation}
\exp\left[i\phi^a\sum_{\mu}^D (E^a_{i,\mu} - E^{(L)}_{i,\mu})\right] \Psi(\{ U\}) = 0 \quad ,
\label{eq:Psitrans}
\end{equation}
where $E^{(L)a}$ are the left-handed generator for $SU(N)$,
\begin{equation}
E^{(L)a}_{\vec i, \mu} U_{\vec i - e_\mu,\vec i} = U_{\vec i - e_\mu,\vec i}
\left( \frac{1}{2} \lambda^a \right) \quad ,
\end{equation}
which give rise to the right-multiplication in Eq.~(\ref{eq:gaussinvb}).

From the infinitesimal from of Eq.~(\ref{eq:Psitrans} we find the lattice form
of the Gauss' law, 
\begin{equation}
\sum_{\mu}^D (E^a_{i,\mu} - E^{(L)}_{i,\mu}) \Psi(\{ U\}) = 0 \quad .
\end{equation}
It is easy to show that
\begin{eqnarray}
E^{(L)a}_{\vec i, \mu} &  = &  D_{ab}(U_{\vec{i}- e_\mu, \vec i} ) 
E^a_{\vec i, \mu} \quad , \\
D_{ab}(U) & = &  \frac{1}{2} \trace \left[ U \lambda^a U^\dagger \lambda^b
\right] \quad , \\
\, [ E^{(L)a}_{\vec i, \mu} , E^{(L)b}_{\vec i, \mu}  ]
& = & - 2 i f^{abc} E^{(L)c}_{\vec i, \mu}  \quad .
\end{eqnarray}
We thus find the lattice version of the Gauss' law generators
\begin{equation}
G^a_{\vec i,\mu} = \left(E^a_{\vec i,\mu} -
 D_{ab}(U_{\vec i- e_\mu ,\vec i})E^a_{\vec i- e_\mu ,\mu} \right) \quad .
\label{eq:latgauss})
\end{equation}
In the limit that the lattice spacing vanishes this simplifies to the
original Gauss' law generator, Eq.~(\ref{gss2}).

From Eq.~(\ref{eq:latgauss}) we see that we can define a covariant
lattice derivative, which takes the matrix form
\begin{equation}
\sunobj D_{\vec i, \vec j, \mu} = \delta_{\vec i+ e_\mu , \vec j} -
 D_{ab}(U_{\vec i- e_\mu ,\vec i}) \delta_{\vec i- e_\mu, \vec j} \quad .
\end{equation}
This construction of the Gauss' law makes it clear why it is so attractive
to work with functions of traces of the product of $U$ along closed loops 
(Wilson loops), since those varaibles are automatically gauge invariant.

\begin{table}
\caption{The number of degrees of freedom on a hypercubic 
lattice of size $n\times n\times \cdots n$, for
different numbers of space dimensions. The $D=1$ case comprises
$n$ plaquettes on a line\label{tab:count}}
\begin{center}
\renewcommand{\arraystretch}{1.2}
\begin{tabular}{l|llll}
dimensionality & ``1'' & 2 & 3  & $D$ \\
\hline 
number of sites & $2(n+1)$ & $(n+1)^2$ &  $(n+1)^3$ & $(n+1)^D$ \\ 
number of links & $3n+1$ & $2n(n+1)$ & $3n(n+1)^2$  & $Dn(n+1)^{D-1}$ \\
 number of plaquettes & $n$ &  $n^2$ & $3n^3$ & $\frac{1}{2} D(D-1) n^D$
\end{tabular}
\end{center}
\end{table}


Thus the number of constraints equals the number of group generators
times the number of sites, and the number of degrees of freedom equals
the number of generators times the number of links. The number of
unconstrained degrees of freedom is clearly the latter number minus
the former. As shown in Table~\ref{tab:count}, for one and two space
dimensions, in the limit of infinite extent, this is exactly the
number of plaquettes times the number of gauge degrees of freedom.
This is not true in higher dimensions, and we shall have to do some
more work to get around this problem.

Another interpretation of this procedure is that Gauss' law fixes the
longitudinal electric field. This field is a direct consequence of the
charge density distribution in the system. Since we will work in the
colourless sector, the longitudinal field is zero, in agreement
with our results above.

\section{Explicit Hamiltonian\label{sec:Ham}}

The link variables are a sufficient set of variables, and, when
combined in closed contours, they are gauge invariant. This means that
projections on the colour-neutral sector by integration over the gauge
group, as seen in many Lagrangian lattice gauge approaches to the
problem, have no effect. The price we have paid for the Hamiltonian
approach is the breaking of explicit Lorentz invariance, and the problem of
determining the physical subspace.  It is also easy to see that the
collection of all possible contour variables form an overcomplete
set. It can be shown explicitly that there are relations between
different combinations of contours, and also between contours winding
a different number of  times around the same  path.

For a systematic Hamiltonian approach, we want to define an exactly
complete (i.e., neither overcomplete nor incomplete) set of variables,
and we must know the effect of the electric and magnetic operators
upon these variables. Part of the overcompleteness is due to the fact
that we still have residual gauge degrees of freedom.  Therefore we
shall fix of all attempt to fix the gauge fully. Some natural choices,
such as Coulomb and maximal Abelian gauges \cite{Sto94,Baa85,Hay98},
are not suited to a Hamiltonian approach. The Coulomb gauge is highly
non-local, which makes localised approximation hard to construct, and
the maximal-Abelian gauge requires a diagonalisation procedure for
each value of the field, not an attractive procedure in an operatorial
approach. The Coulomb gauge was applied in the Hamiltonian framework
by Christ and Lee \cite{CL80}, but that approach has not shed any
light on the properties of QCD.

Inspired by initial work of M\"uller and R\"uhl \cite{MR84}, and
Bronzan \cite{Bro85,Bro88,Bro88b,BV91,BV93}, we shall use a
maximal-tree gauge where we gauge-fix a maximal number of links, which
leaves a complete set of variables in the Hamiltonian.  We combine
these link variables to form closed contours, which is useful in the
absence of colour charges. This is still not enough, since another
form of overcompleteness creeps in related to the number of degrees of
freedom in traces of ${SU}(N)$ matrices.  In the next section,
Sec.~\ref{sec:FC}, we investigate the question of what conditions
exist among different trace variables.

\subsection{Conditions on variables}

As in electromagnetism the magnetic field should be divergenceless. This is
a direct consequence of the skew-symmetry of the field tensor,
\begin{equation}
\sum_{i}^D \sunobj D_{i}(\vec{m})\sunobj B_{i}(\vec{m})=0\quad,
\end{equation}
which follows from the Jacobi identity concerning the vanishing of the
sum over all cyclic permutations of the double commutator $\sum
_{\mathrm{cyclic}}[\sunobj D_{i},[\sunobj D_{j},\sunobj D_{k}]]=0$.
Therefore the magnetic flux through a closed surface should be
zero. Since we have contours and more than one component of the
magnetic field in non-Abelian ${SU}(N)$ gauge theory, similar
relations can be derived for arbitrary contours $\alpha $.  In order
to combine the magnetic flux through a set of contours we define
\cite{Man79}
\begin{eqnarray}
W^0_{\alpha } & \equiv & \trace [U_{\alpha }]\quad,\\
W^{a}_{\alpha } & \equiv & \trace [\lambda ^{a}U_{\alpha }]\quad.
\end{eqnarray}
The set $\{W^0_{\alpha},W^a_{\alpha}\}$
which uniquely determines $U_{\alpha }$, as can be seen from the decomposition
$U_{\alpha }=x^{0}1+x^{a}\lambda ^{a}$.

From commuting the matrices in a trace of arbitrary length we can derive relations
between contours and their decomposition in smaller contours. For example
for $SU(2)$ where $W^0_{\alpha}$ is real, and $W^a_{\alpha}$ is purely 
imaginary we find
\begin{equation}
\trace[U_{\alpha} U_{\beta}] +\trace[U_{\alpha} U^{-1}_{\beta}] 
=\trace[U_{\alpha}]\trace[U_{\beta}] .
\end{equation}
For a general $SU(N)$ we find for the trace of three matrices that
\begin{eqnarray}
\trace[U_{\alpha} U_{\beta} U_{\gamma}]-\trace[U_{\alpha} U_{\gamma} U_{\beta}]
&=& \frac{i}{2} \sum_{a,b,c} f^{abc} 
W_{\alpha}^a W_{\beta}^b W_{\gamma}^c, \\
\trace[U_{\alpha} U_{\beta} U_{\gamma}]+\trace[U_{\alpha} U_{\gamma} U_{\beta}]
&=& 2 \frac{1-3N}{N^2} \trace[U_{\alpha}] \trace[U_{\beta}] \trace[U_{\gamma}]
\nonumber\\
&&+ \frac{1}{N} \left(
\trace[U_{\alpha}]\trace[ U_{\beta} U_{\gamma}]+
\trace[U_{\beta}]\trace[ U_{\alpha} U_{\gamma}]+
\trace[U_{\gamma}]\trace[ U_{\alpha} U_{\beta}]\right)\nonumber\\
&&+\frac{1}{2} \sum_{a,b,c} d^{abc} W_{\alpha}^a W_{\beta}^b W_{\gamma}^c\quad.
\end{eqnarray}
For the case of $SU(2)$ and $SO(3)$ these equations simplify, since in that
case $d^{abc}=0$ and $f^{abc}=\epsilon^{abc}$.
These relations define a set of constraints, usually
called ``Mandelstam constraints'' \cite{Man79}.

Not only do relations exists among contours, but also for paths
winding around the same contour more often than $N-1$ times. These
trace along such a contour can be re-expressed in terms of
the first $N-1$ trace variables,
\begin{equation}
\trace [U^{n}];\quad n=1,2,\cdots ,N-1\quad.
\end{equation}
 So we find 
\begin{equation}
\trace [U^{p}]=\sum _{\xi }c_{\xi }\trace [U]^{\xi _{1}}\trace [U^{2}]^{\xi _{2}}\cdots \trace [U^{N-1}]^{\xi _{N-1}}\quad,
\end{equation}
 where $\sum _{j=1}^{N-1}j\xi _{j}\leq p$. Generally we do not need to
know these relations for all values of $p$. If we determine the eigenstates
of the electric operator the necessary relations follow. We will derive the
values for $c_{\xi }$ for some useful cases in  Appendix A.

For $SU(2)$, which has only one independent trace variable $\trace [U]$,
the relations are, as can be derived from the representation,
\begin{eqnarray}
\trace [U^{-1}] & = & 2 \cos\phi = \trace [U]\quad,\\
\trace [U^{n}] & = & 2 \cos n\phi = 2\sum _{i=0}^{\left[ \frac{n}{2}\right] }
\begin{pmatrix} n\\2i\end{pmatrix}
\left( \frac{\trace [U]}{2}\right) ^{n-2i}\left( 1-\frac{\trace [U]^{2}}{4}\right) ^{i}\quad,
\end{eqnarray}
 and the lowest-order relations for $SU(3)$ are 
\begin{eqnarray}
\trace [U^{-1}] & = & \frac{1}{2}\trace [U]^{2}-\frac{1}{2}\trace [U^{2}]\quad,\\
\trace [U^{3}] & = & 3+\frac{3}{2}\trace [U^{2}]\trace [U]-\frac{1}{2}\trace [U]^{3}\quad.
\end{eqnarray}

\subsection{The gauge-fixed Hamiltonian}

\begin{figure}
\centerline{\includegraphics[width=6cm]{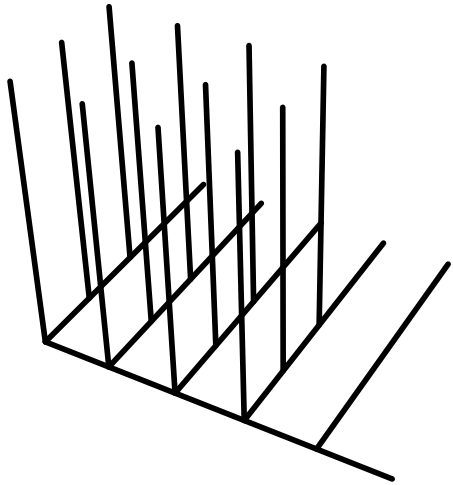}}
\caption{A maximal tree as defined in Eq.~(\ref{eq:deftree}).}
\label{fig:maxtree}
\end{figure}

From the counting of degrees of freedom in Table \ref{tab:count} we know
that the set of all link variables must in general be overcomplete. 
Since the link variables are still one of the most attractive sets to
use, we choose to fix the gauge as much as possible, based on some
ideas developed by M\"uller and R\"uhl \cite{MR84}, and
Bronzan and collaborators
\cite{Bro85,Bro88,Bro88b,BV91,BV93}. The approach is based on fixing the gauge
along the links of a maximal tree, a collection of links on the
lattice which connect any two lattice sites by one and only path.
\begin{equation}
\{U_{\vec{i}\vec{j}}\}\rightarrow \{V_{\vec{n}\vec{l}},W_{\vec{k}\vec{m}}|W\in \mathrm{tree},V\not \in \mathrm{tree}\}\quad .\label{eq:deftree}
\end{equation}

\begin{figure}
\centerline{\includegraphics[width=6cm]{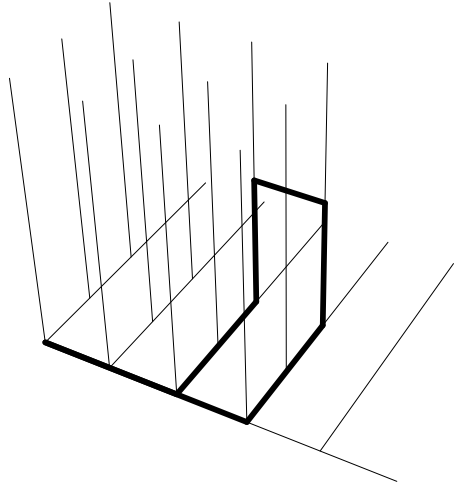}}
\caption{A sketch of the path in a typical variable $X$.}
\label{fig:Xloop}
\end{figure}

These $W$ variables can serve as our ``longitudinal electric fields'',
and are determined by the gauge fixing. The links not on the tree form
the base for the ``magnetic variables''. However, as they stand, these
variables are not invariant under local gauge transformations, which
means that the wavefunction cannot depend directly on these variables,
because that would contradict gauge invariance.  Hence we combine them
with a path over the maximal tree from and to the origin,
\begin{equation}
X_{\vec{i}\vec{j}}=W_{0,\vec{i}} V_{\vec{i}\vec{j}}W^{-1}_{0,\vec{j}} \quad,
\label{eq:defX}
\end{equation}
see Fig.~\ref{fig:Xloop}.  In order to have a unique way of defining
the links on the maximal tree, we have oriented the links on the
maximal tree such that they point along the direction away from the
origin, which is why an inverse appears on the far right in
Eq.~(\ref{eq:defX}), since $W_{0\vec i}$ is defined as the product
along the maximal tree,
\begin{equation}
W_{0\vec{i}}=
\prod _{k=0}^{\gamma ({\vec{i}})-1} W_{\vec{j}_{k}\vec{j}_{k+1}}\quad,
\end{equation}
where $\gamma (\vec{i})$ is the number of links on the maximal tree
between the origin and site $\vec{i}$.  Note that there is no trace in
Eq.~(\ref{eq:defX}), and thus all the variables $X_{\vec{i}\vec{j}}$
transform in the same way under local gauge transformations with the
gauge transformation at the origin, and are invariant under all other
local gauge changes. We know that when we fix the gauge we cannot fix
a global gauge transformation, and we are thus led to identify this
with the one at the origin.

We can both use $X_{\vec{i}\vec{j}}$ and $X_{\vec{j}\vec{i}}$
as variables, but since these are clearly related by an inversion, we
shall just use one of them. In the sum over links in the electric
operator we shall also just use one order, which we choose again to be
such that the link point along a branch of the maximal tree away from
the origin for tree links, and such that the first non-zero component
of the direction of the link is positive for the remaining ones.

The electric operators $E^{a}_{\vec{i}'\vec{j}'}$ defined on every
link naturally fall into two classes: those acting on the maximal tree
variables $W$ and those acting on the variables $V$. Let us first look
at those acting on a tree variable, and evaluate the action on a
variable $X_{\vec i, \vec j}$. Again we distinguish two cases,
since the link can be contained in either (actually even in both!) the
path from $0$ to $\vec i$, or in the path from $\vec j$ to $0$. If the
electric operator acts in the first part we find
\begin{eqnarray}
E^a_{\vec{i}'\vec{j}'} X_{\vec i, \vec j} &=&
\prod_{k=0}^{\gamma(\vec{i}')-1} W_{\vec{j}_k,\vec{j}_{k+1}}
\left(E^a_{\vec{i}'\vec{j}'} W_{\vec{i}'\vec{j}'} \right) 
\prod_{\gamma(\vec{j}')}^{\gamma(\vec i)-1}W_{\vec{j}_k,\vec{j}_{k+1}}
V_{\vec i, \vec j} W_{0,\vec{j}}^{-1}\nonumber\\
&=&
W_{0\vec{i}'} \left(-\frac{1}{2}\lambda^a W_{\vec{i}'\vec{j}'} \right)
W_{\vec{j}'\vec{i}} V_{\vec i, \vec j} W_{0,\vec{j}}^{-1}\nonumber\\
&=& W_{0\vec{i}'} \left(-\frac{1}{2}\lambda^a W_{\vec{i}'\vec{j}'} \right)
W_{0\vec{j}'}^{-1} X_{\vec i, \vec j} \nonumber\\
&=&-\frac{1}{2}W_{0\vec{i}'}\lambda^a W_{0\vec{i}'}^{-1}  X_{\vec i,
\vec j}\quad.
\end{eqnarray}
Similarly, being much less explicit, when acting on the last part of
$X$ we have
\begin{eqnarray}
E^a_{\vec{i}'\vec{j}'} X_{\vec i, \vec j} &=&
\frac{1}{2}  X_{\vec i,\vec j} W_{0\vec{i}'}^{-1}\lambda^a
W_{0\vec{i}'}\quad.
\end{eqnarray}
Finally the action on the non-tree variable will also lead to
left-multiplication
by $\lambda^a$, due to our choice of ordering, and we have
\begin{eqnarray}
E^a_{\vec{i}\vec{j}} X_{\vec i, \vec j} &=&
-\frac{1}{2}W_{0\vec{i}}\lambda^a W_{0\vec{i}}^{-1}  X_{\vec i,\vec j}
\quad.
\end{eqnarray}
These expressions are clearly highly suggestive of defining a new
operator that leads to left or right multiplication by 
$\frac{1}{2}W_{0\vec{i}}\lambda^a W_{0\vec{i}}^{-1}$. Using some simple
elements of representation theory, we actually know that
\begin{equation}
W^{-1}\lambda^a W = \sum_b D^{ab}(W) \lambda^b\quad,
\end{equation}
with 
\begin{equation}
D^{ab}(W) = \frac{1}{2} \trace [W^{-1} \lambda^a W \lambda^b]\quad.
\end{equation}
If we define a set of ``intrinsic'' electric operators by
\begin{equation}
\mathcal{E}^a_{\vec i \vec j} \equiv D^{ab}(W_{0\vec{i}}) E^b_{\vec i \vec
j}\quad,
\end{equation}
it is a matter of straightforward algebra to show that these new
operators act much more simply on $X$, leading to a left
multiplication by $-\frac{1}{2}\lambda^a$ if the labels on the link
referred to by the electric operator occur in the same order in the
product, and a right multiplication by $\frac{1}{2}\lambda^a$ if it
appears in inverse order. Now we make use of the unitarity of the
representation functions,
\begin{equation}
\sum_{a=1}^{N^2-1} D_{ab}^* D_{ac} = \delta_{bc}\quad,
\end{equation}
to show that the square of the electric operator, which is the
expression that occurs in the Hamiltonian, is invariant under the
transformation, 
\begin{equation}
\sum_a \left(E^a_{\vec i \vec j}\right)^2
= \sum_a \left(\mathcal{E}^a_{\vec i \vec j}\right)^2\quad.
\end{equation}

If we assume that the wave function can be expressed in terms of the
variables $X_{\vec i \vec j}$, we need to find the action of the sum
over the whole lattice of the electric energy, proportional to the
square of the electric field, on this function.  Since there are two
electric operators in the square, they can either both act on the same
variable, or on different variables. In the first case we find
contributions where $X_{\vec{i}\vec{j}}$ is multiplied from the left
with $-\frac{1}{2}\lambda^a$, for all electric operators along the path leading up
to the site $\vec{i}$, as well as the link $\nnb{ \vec{i}\vec{j}} $ itself,
and multiplied from the right with $\frac{1}{2}\lambda^a$ 
back from the site $\vec{j}$ to the origin,
\begin{equation}
\sum _{\nnb{\vec{i}'\vec{j}'}}
\left(\mathcal{E}_{\vec{i}'\vec{j}'}^{a}\right)^2
X_{\vec{i}\vec{j}}=
-\frac{1}{2}(\gamma (\vec{i})+1)\lambda ^{a}X_{\vec{i}\vec{j}}+
\frac{1}{2}\gamma (\vec{j})X_{\vec{i}\vec{j}}\lambda ^{a}\quad.
\end{equation}
In the second case each electric operator acts on a different variable
$X$,
\begin{eqnarray}
\sum _{\nnb{\vec{i}'\vec{j}'}}
\left[ \mathcal{E}_{\vec{i}'\vec{j}'}^{a} X_{\vec{i}\vec{j}}\right]
\left[ \mathcal{E}_{\vec{i}'\vec{j}'}^{a} X_{\vec{k}\vec{l}}\right]
 & = & 
\frac{1}{4}\left(\gamma (\vec{i}\cap \vec{l}) +
	\delta_{\nnb{\vec{i}\vec{j}},\nnb{\vec{k}\vec{l}}}\right)
\lambda ^{a}X_{\vec{i}\vec{j}}\lambda ^{a}X_{\vec{l}\vec{m}}
\nonumber\\
 & & - \frac{1}{4}\gamma (\vec{i}\cap \vec{m})
	\lambda ^{a}X_{\vec{i}\vec{j}}X_{\vec{l}\vec{m}}\lambda ^{a}
\nonumber\\
 & & - \frac{1}{4}\gamma (\vec{j}\cap \vec{l})
	X_{\vec{i}\vec{j}}\lambda ^{a}\lambda ^{a}X_{\vec{l}\vec{m}}
\nonumber\\
 & & + \frac{1}{4}\gamma (\vec{j}\cap \vec{m})
X_{\vec{i}\vec{j}}\lambda ^{a}X_{\vec{l}\vec{m}}\lambda ^{a}\quad,
\end{eqnarray}
 where the notation $\gamma (\vec{i}\cap \vec{j})$ denotes the number of
links common between the maximal-tree paths from 0 to $\vec{i}$ and from
0 to $\vec{j}$.

We now wish to show in what sense the variables $X$ are sufficient.
Given a general wave function in terms of both the $X$ and the $W$
variables we investigate when this wave function lies in the colour-neutral
sector.
Then it  has to satisfy Gauss' law,
\begin{equation}
{\mathcal{G}}^{a}(\vec{i})|\psi \rangle =0;\quad\, \, \forall {\vec{i}}\not =0\quad,
\end{equation}
which can be reformulated as a local gauge transformation generated by Gauss'
law,
\begin{equation}
e^{i\lambda ^{a}{\mathcal{G}}^{a}(\vec{i})}|\psi \rangle =|^{\lambda ^{a}{\mathcal{G}}^{a}}\psi \rangle =|\psi \rangle \quad.
\end{equation}
In a $\{X,W\}$ representation the wave function changes to 
\begin{equation}
\langle \{X_{\vec{k}\vec{j}},W_{\vec{l}\vec{m}}\}|^{\sunobj{\phi }}\psi \rangle =
\langle \{\sunobj{\phi}^{-1}(0)X_{\vec{k}\vec{j}}\sunobj{\phi}(0),\sunobj{\phi}^{-1}(\vec{l})W_{\vec{l}\vec{m}}\sunobj{\phi} (\vec{m})\}|\psi \rangle \quad.
\end{equation}
We have already identified the transformation $\sunobj \phi(0)$ as the global
gauge transformation, and we are not concerned about non-invariance
under these transformations. The $W$ variables are clearly not
invariant under local gauge transformations, and a function in the
physical subspace is thus seen to be a function of the $X$  variables
only. 

This approach can also be identified with a gauge fixing where we set
the matrices on the maximal tree links to unity,
\begin{equation}
\{V_{\vec{i}\vec{j}},W_{\vec{k}\vec{l}}\}\rightarrow\{X_{\vec{i}\vec{j}},1\}\quad,
\end{equation}
but we keep track of the fact that the electric fields on the
gauge-fixed links still act non-trivially on the physical variables.

Let us now study the effect of the global gauge transformation
$\sunobj{\phi}(0)$,
\begin{equation}
|^{\sunobj \phi (0)}\psi \rangle =U(\sunobj {\phi })|\psi \rangle\quad.
\end{equation}  
We know that a colourless wave function transforms as a singlet under
global gauge transformations (i.e., it is invariant under
$\sunobj{\phi}$).  One way to make sure that the wave function
satisfies this condition is to combine the $X$ variables in the wave
function in traces, each of which  transforms as a colour singlet,
\begin{equation}
\trace [X^{p}_{\alpha }],\quad\, \, \trace [X^{p}_{\alpha }X^{q}_{\beta }],\quad\, \, \trace [X^{p}_{\alpha }X^{q}_{\beta }X^{r}_{\gamma }],\quad\cdots .
\end{equation}
However, as we have seen before, this set is overcomplete; there are
constraints among the different trace variables. In another approach
to this problem we can relate the variables to group elements in
colour space, and in the colourless sector the wave function can
depend on all scalars that can be defined with these group
elements. If we use the Mandelstam variables \cite{Man79}
\begin{equation}
W_{\alpha }=\trace [X_{\alpha }];\quad\, \, W^{a}_{\alpha }=\trace [\lambda ^{a}X_{\alpha }]\quad\, \, ,
\end{equation}
we find that there must be a constraint among them, as there are only
$N^{2}-1$ independent degrees of freedom. Any unitary matrix
$V=(\vec{v}_{1},\vec{v}_{2},\cdots ,\vec{v}_{N})\in {SU}(N)$ has
$2N^{2}$ degrees of freedom satisfying $N^{2}+1$ constraints ($\det
V=1$, $\overline{\vec{v}}_{i}\cdot \vec{v}_{j}=\delta_{ij}$).
Although the number of degrees of freedom equals  the number
$W^{a}_{\alpha }$'s, the constraint do not fully fix the additional
degree of freedom $(W_{\alpha })$, which can take on several discrete
values, corresponding to discrete gauge transformations.  For example,
$SU(2)$ can be represented by a unit vector on the four-dimensional
sphere. In this case the discrete gauge transformation is the sign of
$W_{\alpha}^0$, which determines whether the vector is on the positive
or the negative hemisphere. We shall study an approach to these
problems in the next section.

We shall close this section by making a particular choice of maximal
tree, and defining the Hamiltonian within the physical subspace
explicitly, in a recursive way. First of all (for $D=3$) we choose all
links on the $x$-axis for $y=z=0$, all the links in the $y$-direction for
$z=0$, and all links in the $z$ direction, as in Fig.~\ref{fig:maxtree}.
In a more mathematical notation the maximal tree can be defined as the
union
\begin{equation}
\label{maxtree}
\bigcup \{[(i_{x},0,0)(i_{x}+1,0,0)];
          [(i_{x},i_{y},0)(i_{x},i_{y}+1,0)];
          [(i_{x},i_{y},i_{z})(i_{x},i_{y},i_{z}+1)]\}\quad.
\end{equation}
The overlap of two paths is given by
\begin{equation}
\gamma (\vec{i}\cap \vec{j})=\min (i_{x},j_{x})+
\delta _{i_{x}j_{x}}\min (i_{y},j_{y})+\delta _{i_{x}j_{x}}\delta _{i_{y}j_{y}}\min (i_{z},j_{z})\quad.
\end{equation}

We start from the ($D=3$) Kogut-Susskind Hamiltonian,
Eq.~(\ref{eq:HKS}),
and let it act on a function of the $X$ variables. For the electric
part we have
\begin{eqnarray}
H_{E} & = & 
\frac{1}{4}\sum _{\nnb{ \vec{k}\vec{l}} ,\nnb{\vec{i}\vec{j}}\not\in\text{tree}} \sum_{a}\Bigg \{
\gamma (\vec{k}\cap \vec{i})
	\left[ \lambda ^{a}X_{\vec{k}\vec{l}}
		\frac{\delta }{\delta X_{\vec{k}\vec{l}}}\right] 
	\left[ \lambda ^{a}X_{\vec{i}\vec{j}}
		\frac{\delta }{\delta X_{\vec{i}\vec{j}}}\right] 
\nonumber \\
 &  & \qquad
-\gamma (\vec{l}\cap \vec{i})
	\left[ X_{\vec{k}\vec{l}}\lambda ^{a}
		\frac{\delta }{\delta X_{\vec{k}\vec{l}}}\right] 
	\left[ \lambda ^{a}X_{\vec{i}\vec{j}}
		\frac{\delta }{\delta X_{\vec{i}\vec{j}}}\right] 
\nonumber \\
 &  & \qquad
-\gamma (\vec{k}\cap \vec{j})
	\left[ \lambda ^{a}X_{\vec{k}\vec{l}}
		\frac{\delta }{\delta X_{\vec{k}\vec{l}}}\right] 
	\left[ X_{\vec{i}\vec{j}}\lambda ^{a}
		\frac{\delta }{\delta X_{\vec{i}\vec{j}}}\right] 
\nonumber\\
 &  & \qquad
+\gamma (\vec{l}\cap \vec{j})
	\left[ X_{\vec{k}\vec{l}}\lambda ^{a}
		\frac{\delta }{\delta X_{\vec{k}\vec{l}}}\right] 
	\left[ X_{\vec{i}\vec{j}}\lambda ^{a}
		\frac{\delta }{\delta X_{\vec{i}\vec{j}}}\right] \Bigg
\}
\nonumber\\&&
\qquad
+\frac{1}{4}\sum _{\nnb{ \vec{i}\vec{j}} ,a}
	\left[ \lambda ^{a}X_{\vec{i}\vec{j}}
		\frac{\delta }{\delta X_{\vec{i}\vec{j}}}\right] 
	\left[ \lambda ^{a}X_{\vec{i}\vec{j}}
		\frac{\delta }{\delta X_{\vec{i}\vec{j}}}\right] \quad,
\end{eqnarray}
where the functional derivative combined with the matrix $\lambda^a$
is meant to denote the insertion of $\lambda ^{a}$ at a position where
the variable occurs, e.g.,
\begin{equation}
\left[ \lambda ^{a}X_{\vec{i}\vec{j}}\frac{\delta }{\delta
X_{\vec{i}\vec{j}}}\right] 
f(X^{p}_{\vec{i}'\vec{j}'})=
\delta_{\vec{i}\vec{i}'} \delta_{\vec{j}\vec{j}'} 
\left( f'(\lambda^{a} X^{p}_{\vec{i}\vec{j}})+
f'(X_{\vec{i}\vec{j}}\lambda^{a}X^{p-1}_{\vec{i}\vec{j}})+\cdots
+f'(X^{p-1}_{\vec{i}\vec{j}}\lambda ^{a}X_{\vec{i}\vec{j}})\right)
\quad.
\end{equation}

In the magnetic part we can  set all of the links on the maximal tree to
unity, 
\begin{eqnarray}
H_{M} & = &
\sum _{\text{plaquette}}\trace 
\left[2-U_{\mathrm{plaquette}}-U^{-1}_{\mathrm{plaquette}}\right] 
_{U_{\mathrm{tree}}\rightarrow 1}\nonumber\\
& = & \sum _{i}\trace
	\left[2-X_{(i,1,0)(i+1,1,0)}-X^{-1}_{(i,1,0)(i+1,1,0)}\right]
\nonumber\\ &  & 
+\sum_{i,j\geq1}\trace\left[2 -
	X_{(i,j,0)(i+1,j,0)}X^{-1}_{(i,j+1,0)(i+1,j+1,0)} -
	X_{(i,j+1,0)(i+1,j+1,0)}X^{-1}_{(i,j,0)(i+1,j,0)}\right] 
\nonumber\\ &&  +
\sum _{i,j,k}\trace\left[2 -
X_{(i,j,k)(i+1,j,k)}X^{-1}_{(i,j,k+1)(i+1,j,k+1)} -
X_{(i,j,k+1)(i+1,j,k+1)}X^{-1}_{(i,j,k)(i+1,j,k)}\right] 
\nonumber\\ &  &+
\sum _{i}\trace
	\left[2-X_{(i,0,1)(i+1,0,1)}-X^{-1}_{(i,0,1)(i+1,0,1)}\right]
\nonumber\\ &  &
 +\sum _{\substack{i,j,k\\ (j,k)\neq(0,1)}}
\trace\left[2 -
X_{(i,j,k)(i,j+1,k)}X^{-1}_{(i,j,k+1)(i,j+1,k+1)} -
X_{(i,j,k+1)(i,j+1,k+1)}X^{-1}_{(i,j,k)(i,j+1,k)}\right] 
\nonumber\\ & &+ 
\sum _{i,j}\trace
	\left[2-X_{(i,j,1)(i,j+1,1)}-X^{-1}_{(i,j,1)(i+1,j+1,1)}\right]
\nonumber\\ &  &+ 
\sum _{i,j,k}\trace\left[2 - 
	X_{(i,j,k)(i+1,j,k)}X_{(i+1,j,k)(i+1,j+1,k)}
	X^{-1}_{(i,j+1,k)(i+1,j+1,k)}X^{-1}_{(i,j,k)(i,j+1,k)} \right.
\nonumber\\ &  &
\qquad \left.-X_{(i,j,k)(i,j+1,k)}X_{(i,j+1,k)(i+1,j+1,k)}
	X^{-1}_{(i+1,j,k)(i+1,j+1,k)}X^{-1}_{(i,j,k)(i+1,j,k)}\right]\quad .
\label{eq:MagFixed}
\end{eqnarray}
 The first sum corresponds to plaquettes is the $xy$-plane, touching the
$x$-axis, which contain only one link not on the maximal tree. The second
sum contain all the other links in the $xy$-plane. The third and fourth
sum contain plaquettes in constant-$y$ planes, 
and the fifth and sixth in  the constant-$x$ planes,
respectively. The last sum contains the plaquettes in the $z$-planes for
$z \not = 0$.

\subsection{Boundaries}
Let us spend a little time counting degrees of freedom, to see how
close we are to the expected number.  On a finite lattice of length
$n$ in $D$ dimensions there are $(n+1)^{D}$ lattice points, and
$D(n+1)^{D-1}n$ links. The maximal tree contains $n +
n(n+1)+n(n+1)^2+\ldots+n (n+1)^{D-1}n=(n+1)^D-1$ links, as can easily
be seen from (the $D$-dimensional generalisation of) our explicit
choice of maximal tree. Therefore there are $(D-1)N^{D}-N^{D-1}-\cdots
-N$ remaining links, and a similar number of variables $X$. Each of
these variables has $N^2-1$ degrees of freedom. The corresponding
total number of freedom is thus just $N^2-1$ bigger than the number of
unconstrained degrees of freedom discussed in
Sec.~\ref{sec:Gauss}. The additional degrees of freedom are associated
with the global gauge degree of freedom, represented by the gauge
freedom at the origin.

\begin{table}
\caption{The number of degrees of freedom in the maximal tree approach
lattice of size $n\times n\times \cdots n$, for
different number of space dimensions.\label{tab:count2}}
\begin{center}
\renewcommand{\arraystretch}{1.2}
{\scriptsize
\begin{tabular}{l|lll}
dimensionality & 2 & 3 & $D$  \\ \hline 
number of sites & $(n+1)^2$ & $(n+1)^3$ & $(n+1)^D$ \\
number of links & $2 n(n+1)$  & $3 n(n+1)^2$  & $Dn(n+1)^{D-1}$ \\
number of links on tree&  $n(n+1)+n$ & $3(n+1)^2-1$ &
$(n+1)^D -1$ \\
 number of independent variables &$n^2$ &$2n^3+3n^2$ &
$(D-1)  (n+1)^{D} -D(n+1)^{D-1}+1$
\end{tabular}}
\end{center}
\end{table}

Our method shows some similarity to attempts to quantise QCD on a
torus \cite{LenzNausThies}. On a torus we have to impose periodic
boundary conditions, where identify the $n+1$-st point on the lattice
with the first point. The same maximal tree applies, however, we
cannot fix the fields on the links from the $n$th point to the $n+1$st
point, since in that case we would have a closed path, and would
violate the uniqueness of paths on the maximal tree. Thus we have an
additional $D(n+1)^{D-1}$ free links connecting the boundaries of the
lattice. Therefore there are $D(n+1)^{D-1}$ extra independent
variables $X$. $D$ of these new degrees of freedom are Polyakov loops,
which wind around the lattice \cite{Wat94}.

One of the goals of our work is to apply many-body methods to the
Hamiltonian, in all likelihood eventually we will apply the coupled
cluster method, which expresses the wave function in terms of
correlations between different variables. For an efficient
implementation, we assume translational invariance, which restricts
the number of independent coefficients in the wave functional
considerably. Correlations on a periodic lattice are truly
translationally invariant. However, on a finite lattice, if we take the
lattice size to infinity the system should have a inner region where
translational invariance is observed. In practice we assume an
infinite lattice.

\section{Further constraints}\label{sec:FC}
Even though we have drastically reduced the number of degrees of
freedom, additional complications arise when we impose
colour neutrality on the wave function. Here, the natural choice is to
work with traces of products of the variables $X$, as discussed above.
A suitable approach would be to construct a basis of eigenstates of
the electric part of the Hamiltonian and calculate matrix elements of
the magnetic energy between these states. Such an approach is a quite
natural calculational scheme for the Hamiltonian approach. One can also
use the method inherent in the Lagrangian calculations, based on
invariant integration over the full group \cite{HK84}, using Monte Carlo
integration. However, for a proper Hamiltonian approach this discards
many of the advantages of the method.

To find eigenstates of the electric operator, one can resort to three
general approaches. Firstly, group theory gives us, in principle, a
way to construct general eigenstates, the group characters.  However,
for a large basis, and $N>2$, this is hopelessly involved \cite{BR65},
unless it can be automated, and we see no easy way to do this.

A second approach is based on integrating configurations, and
constructing orthogonal combinations from them. In this case one must start
off with  much larger overcomplete sets of configurations and at
increasing orders the integration, generally based on Creutz's
integration method \cite{Cre78,Cre83}, tends to be more and more involved
\cite{Leo98}.

The third approach is based on the action of the electric operator
itself, which leads to a block-diagonal matrix which has to be
diagonalised to recover the eigenstates. In combination with a
symbolic method, this seems to be the most powerful approach, which
allows one to tackle any arbitrary $SU(N)$ group. We shall develop
this method here. The method also applies to schemes other than the maximal
tree method, and we shall therefore discuss it in a slightly
more general framework (see also Refs.~\cite{Schuette,bgl89}).

We shall look at states in the colourless sector, where all variables
can be formulated in terms of traces.  Clearly only (products of)
traces defined over the same set of links can mix under the action of
the electric operator (since states with differing links are trivially
orthogonal).  Within such a choice the full set of eigenstates separates into
 families which are characterised by the number of lines along
each link.  In the case of a single plaquette we have the collections:
$\{1\}$, $\{\trace[U]\} $, $\{\trace[U]^2,\trace[U^2]\}$, $
\{\trace[U^3],\trace[U^2]\trace[U],\trace [U]^3 \} , \cdots $. 
Within each collection, we can form linear
combinations of the elements that are  eigenstates of the
electric operator. However, since for $SU(N)$ the contour ${\rm Tr}[U^p]$
with the winding number $p\geq N$ can be expressed in terms of contours with
lower winding numbers, one finds, as one re-expresses these contours
with higher winding numbers, that some eigenstates do not contain
leading order terms, but reduce to states of other, lower-order,
collections.

The symbolic approach starts  by writing explicit indices on
all the traces, labelling the column indices $i_1,\ldots,i_n$. Since
every member function has no free indices, the set of row indices must
be equal to the set of column indices, but the order in which they
appear can clearly be different. Actually, all the other members in
the collection correspond to a distinct permutation, but not every
permutation generates a new member of the collection.  For example a
segment with three lines has six permutations:
\begin{equation}
(123),\quad(213),\quad(132),\quad (321) ,\quad (231),\quad(312)\quad ,
\end{equation}
where, in the case of a single loop, when one end of the
segment is contracted with the other, the first element corresponds to
${\rm Tr}[U]^3$, the next three elements to ${\rm Tr}[U^2] {\rm Tr}[U]$,
and the last two elements to ${\rm Tr}[U^3]$. This follows easily if
one considers the general form:
\begin{equation}
(P_1,P_2,P_3) = \sum_{i_1,i_2,i_3} U_{i_1 i_{P_1}} U_{i_2 i_{P_2}} U_{i_3 i_{P_3}}\quad,
\end{equation}
where $P=(P_1,P_2,P_3)$ is a permutation of $(i_1,i_2,i_3)$.
However, in the case that two lines are connected in one loop $W$
 and the last line is connected to a separate single-line 
loop $V$, i.e. $\sum (WU) (WU) (VU)$, all the permutations are distinct.

If we wish to evaluate the action of the electric part of the Hamiltonian
on such a segment $(i,j,k)$, we first need to understand what
happens if we look at a case where each electric operator acts on a different
matrix $U$,
\begin{eqnarray}
\frac{1}{4}\sum_a (\lambda^a U)_{i_1,i_{P_1}} 
	(\lambda^a U)_{i_2,i_{P_2}}
& = &
\frac{1}{2}  U_{i_2,i_{P_1}} 	U_{i_1,i_{P_2}}
-\frac{1}{2N} U_{i_1,i_{P_1}} 	U_{i_2,i_{P_2}} \nonumber \\
& = & \frac{1}{2}  U_{i_1,i_{P_2}} 	U_{i_2,i_{P_1}}
-\frac{1}{2N} U_{i_1,i_{P_1}} 	U_{i_2,i_{P_2}}\quad.
\label{eq:EasPermu}
\end{eqnarray}
In the last step we have just changed the order in which we  write
the $U$ matrix elements, suggesting that we can interpret this part of
the action of the electric fields as the
interchange of $P_1$ and $P_2$. The result Eq.~(\ref{eq:EasPermu}) is
based on a standard relation for the $SU(N)$ generators $\lambda^a$,
\begin{equation}
\sum_{a=1}^{N^2-1} \lambda^a_{ij} \lambda^a_{kl} =
2 \delta_{il}\delta_{jk}
-\frac{2}{N}\delta_{ij}\delta_{kl}\quad.
\end{equation}
Using the relation (\ref{eq:EasPermu}) we find that the only
non-trivial effect of the electric part of the Hamiltonian is an
exchange of pairs,
\begin{equation}
\sum_{\text{links}} E^2 (i,j,k) = \frac{L}{2}[(j,i,k) + (i,k,j) + (k,j,i)
] - \frac{L}{2N}\left(\left( \begin{matrix} n \cr 2 \end{matrix}
\right) + n (N^2-1) \right) (i,j,k)\quad,
\end{equation}
where $L$ is the length of the segment (i.e., the number of electric
operators acting on it) and $n=3$ is the number of lines.
Note that the electric operator for the different $SU(N)$ groups
is identical up to a multiple of the identity operator, 
so it will lead to the same diagrams for the eigenstates. Differences 
follow from the fact that the point where states with higher winding
numbers are no longer independent depends on $N$.
Considering again the example discussed above of a three-line one-loop 
diagram for $SU(2)$ we find the states, by diagonalising the electric operator,
\begin{eqnarray}
\phi_1 & = & (123)+(213)+(132)+(321)+(231)+(312)\quad, \nonumber\\
\phi_2 & = & (123)-(213)-(132)-(321)+(231)+(312) \quad, \nonumber\\
\phi_3 & = & (123)-(312) \quad, \nonumber\\
\phi_4 & = & (123)-(231) \quad, \nonumber\\
\phi_5 & = & (213)-(321) \quad, \nonumber\\
\phi_6 & = & (213)-(132) \quad.
\end{eqnarray}

In the one-loop case $\phi_5$ and $\phi_6$ vanish, and $\phi_3$ and 
$\phi_4$ are identical. Explicitly, we find
\begin{eqnarray}
\phi_1 & = & 6 C^1_3\left(\frac{1}{2}\trace U \right) \quad, \\
\phi_2 & = & 0 \quad, \\
\phi_3 & = & \phi_4 =  3 C^1_1\left(\frac{1}{2} \trace U \right) \quad, \\
\phi_5 & = & \phi_6  =  0 \quad.
\end{eqnarray}
Here $C^i_j$ are Gegenbauer polynomials \cite{MOS66}.
The state labelled $\phi_1$ is actually the third eigenstate of
the Casimir operator of  $SU(2)$. In general the $n$th eigenstate for $SU(2)$ 
follows from the symmetric combination of the permutations of $n$ elements,
as other combinations, which contain traces of higher powers of $U$,
collapse to lower eigenvalue states. In the example above $\phi_3$
reproduces the first eigenstate.

In this way we can find eigenstates for fully linked traces (i.e.,
where the paths have a link in common); we can obviously use these
results to get eigenstates for unlinked traces by multiplying
eigenstates of the separate segments. The eigenvalue corresponding to
such eigenstates is the sum of the eigenvalues of each segment times
their length. We also have not yet considered expressions containing
both the matrices $U$ and $U^\dagger$.  In that case the collection
also contains elements where the indices of $U$'s and $U^\dagger$'s
are contracted with each other, which leads to a cancellation.
However, this should be a straightforward extension.

\section{The one-plaquette problem\label{sec:OP}}

The above approach is designed for application with a many-body method
such as the coupled-cluster method \cite{CCMpapers}. Before doing so
we wish to consider the simplest limit, where all plaquettes occur
independently \cite{Het93,Hal93}. Whether these plaquettes contain
one, two, or four $X$ variables, Eq.~(\ref{eq:MagFixed}), the
plaquette variable is the only relevant combination. Other variables
are trivially integrated over, which in principle requires knowledge
of the group-invariant Haar measure.  For the purposes of the present
paper knowing the existence of such a measure is sufficient, and all
but one of the $X$ variables around the plaquette are integrated
over. This last remaining matrix we call $U$. The wave function
depends on the trace of powers of the plaquette variable only, which
can be re-expressed in terms of class label, a generalisation of the
group characters $\chi (U)$. It is slightly simpler to investigate the
${U}(N)$ problem first; the ${SU}(N)$ case can be simply
embedded by imposing the constraint $\det (U)=1$.

We wish to analyse the Hamiltonian in various ways, both in terms of the trace
variables
\[\xi _{p}=\frac{1}{p}\trace (U^{p}),\]
(we divide by $p$ for future convenience) and in term of the
eigenvalues of the matrix $U$, which are $N$ unimodular numbers of the
form $e^{i\psi _{i}}$.  The latter parameters are often used in a
group-theoretical context, see in particular Weyl's seminal work
\cite{Weyl}. The most complicated part of the problem arises in the
evaluation of the action of the electric operators, or, in group
theoretical terms, the quadratic Casimir operator, on a function of
either the $\xi $'s or $\psi $'s. The potential is rather trivial
being the sum of traces of $U$ and $U^{\dagger }$.

If we start with a function $F$ of the $\xi $'s, we find that the
action of the electric part of the Hamiltonian can be expressed in terms of
these trace variables as
\begin{eqnarray}
E^{2}_{l}F\left( \{\xi \}\right)  & = & \sum _{p,q}\left
( \frac{1}{2}\trace [U^{p+q}]\right) F_{,pq}
\nonumber\\
 &  &+ \sum _{p}\left( \frac{N^{2}}{2N}\trace [U^{p}] + 
\frac{1}{2}\sum _{q=1}^{p-1}\trace [U^{p-q}]\trace [U^{q}]\right)
F_{,p}
\quad,\label{E2} 
\end{eqnarray}
which can also be generalised to multi-plaquette problems. (Here the
notation $F_{,p}$ denotes the derivative with respect to $\xi_p$, and
multiple indices denote multiple derivates.)  This relation follows
from an identity for the generators of $U(N)$,
\begin{equation}
\sum _{a=1}^{N^{2}}\lambda ^{a}_{ij}\lambda ^{a}_{kl}
	=2\delta _{il}\delta _{jk}\quad.
\end{equation}
The first part of the right-hand side of the expression (\ref{E2}) is
in principle a complicated function of the $\xi $'s if $p+q>N$, and we
see that the action of the Hamiltonian will be highly non-trivial in
these variables.  The same holds for the trace of the inverse of $U$
contained in the potential.

Thus solving the one-plaquette problem in this way is difficult, since
it involves re-expressing $\trace [U^{p}]$ where $p<0$ or $p>N$ in the
original trace variables. We shall show that it is easier to use the
eigenvalue variables, and to perform at the same time a simple
transformation which maps a problem which is symmetric under the
interchange of the eigenvalues (i.e., a bosonic problem) onto a
fermionic problem, as discussed in the next section. This is similar to,
albeit slightly more complicated than, what happens for Yang-Mills
theories on a circle \cite{YMcirc}.

\subsection{Angular representation}

The trace of a  matrix is invariant under unitary transformations,
$\trace [U^{n}]=\trace [(VUV^{-1})^{n}]$.
Therefore we can diagonalise the unitary matrix $U$ in the one-plaquette
problem. From the $N^{2}$ degrees of freedom for $U(N)$ only $N$
are relevant to us, 
\begin{equation}
U=\diag \left( e^{i\psi _{1}},e^{i\psi _{2}},\ldots ,e^{i\psi _{N-1}},e^{i\psi _{N}}\right)\quad,
\label{eq:Ueigval}
\end{equation}
where for ${SU}(N)$, $\sum \psi _{i}=0$, since $\det U=1$. The eigenvalues
$\exp \{i\psi _{i}\}$ of the unitary matrix $U$ lie on the complex
unit circle, as easily follows from the properties of Hermitian matrices by
the Cayley transformation. 

We use Eq.~(\ref{E2}), and expand the trace variables in terms of eigenvalues,
\begin{equation}
\xi _{p}=\frac{1}{p}\sum _{i=1}^{N}e^{ip\psi _{i}}\quad ,
\end{equation}
 and assume $f(\{\xi \})=g(\{\psi \})$. It now becomes clear why we have
absorbed the factor $1/p$ in the definition of the trace variables, the
Jacobian matrix is of the Vandermonde form
\begin{equation}
\frac{\partial \xi _{p}}{\partial \psi _{i}}\equiv J_{pi}=ie^{ip\psi
_{i}}\quad .
\end{equation}
This allows us to re-express the action of the electric energy on a
general function $g(\{\psi\})$, 
\begin{equation}
E^{2}g  =  \left( (J^{-1})_{ip}g_{,i}\right) _{,j}(J^{-1})_{jp}
\end{equation}
After some formal manipulations involving Vandermonde matrices, which
are discussed in the appendix, we find the simple result
\begin{equation}
E^{a}E^{a}g(\{\psi \})=-
\frac{1}{2}\sum _{i}\frac{1}{\Xi ^{2}}
\frac{\partial }{\partial \psi _{i}}
\left( \Xi ^{2}\frac{\partial }{\partial \psi _{i}}g(\{\psi \})\right)
\quad,
\end{equation}
a Laplace-Beltrami operator in a curvilinear coordinate system. The quantity
$\Xi $ is proportional to a slightly generalised Vandermonde determinant,
\begin{eqnarray}
\Xi  & \equiv & (-i)^{N(N-1)/2}\left| \begin{array}{ccccc}
e^{i(N-1)\psi _{1}/2} & e^{i(N-3)\psi _{1}/2} & \ldots  
	& e^{-i(N-3)\psi _{1}/2}& e^{-i(N-1)\psi _{1}/2}\\
e^{i(N-1)\psi _{2}/2} & e^{i(N-3)\psi _{2}/2} & \ldots  
	& e^{-i(N-3)\psi _{2}/2}& e^{-i(N-1)\psi _{2}/2}\\
\vdots  & \vdots  & \ddots  & \vdots & \vdots \\
e^{i(N-1)\psi _{N}/2} & e^{i(N-3)\psi _{N}/2} & \ldots  
	& e^{-i(N-3)\psi _{N}/2}& e^{-i(N-1)\psi _{N}/2}\\
\end{array}\right| \nonumber\\
 & = & (-i)^{N(N-1)/2}\prod _{i}e^{-i(N-1)\psi _{i}/2}\prod _{i<j}\left( e^{i\psi _{i}}-e^{i\psi _{j}}\right) \nonumber\\
 & = & 2^\frac{N(N-1)}{2}\prod _{i<j}\sin\left( (\psi _{i}-\psi _{j})/2\right)\quad .
\label{eq:defXi}
\end{eqnarray}
The intermediate expression masks the fact that $\Xi $ is a real
quantity, which is more obvious from the first and last
expressions. In related group-theoretical
literature, especially those following Weyl's original analysis
\cite{Weyl} one often meets the quantity
\begin{equation}
\Delta =\left| \begin{array}{ccccc}
e^{i(N-1)\psi _{1}} & e^{i(N-2)\psi _{1}} & e^{i(N-3)\psi _{1}} & \ldots  & 1\\
e^{i(N-1)\psi _{2}} & e^{i(N-2)\psi _{2}} & e^{i(N-3)\psi _{2}} & \ldots  & 1\\
\vdots  & \vdots  & \vdots  & \ddots  & \vdots \\
e^{i(N-1)\psi _{N}} & e^{i(N-2)\psi _{N}} & e^{i(N-3)\psi _{N}} & \ldots  & 1\\
\end{array}\right| =\prod _{i<j}(e^{i\psi _{i}}-e^{i\psi _{j}})
\quad,
\end{equation}
rather than $\Xi $. These are simply related by
\begin{equation}
\Xi = (-i)^{N(N-1)/2} \exp\left(-i(N-1)/2\sum_i \psi_i\right)\Delta \quad.
\end{equation}

The Jacobian $J=\Xi ^{2}$ vanishes if two eigenvalues $x_{i}=\exp
i\psi _{i}$ are identical. The Jacobian is antisymmetric under the
permutation of eigenvalues, which corresponds to the reflection
symmetry on the manifold. For $U(2)$ we can analyse these properties
in full. A unitary matrix $U\in U(2)$ can be represented as
$U=e^{i\Psi}(n^{0}+i\vec{n}\cdot \vec{\sigma })$, with $(n_0,\vec n)$
real and $n_0^2+\vec{n}^2=1$. The real phase $\Psi$ is arbitrary, but
is given explicitly as $\Psi = -i\ln[{\rm det}U]/N$.  A simple
diagonalisation shows that this has the two eigenvalues $x_{\pm
}=e^{i\Psi}(n^{0}\pm i|\vec{n}|)$, which are identical for points on
both poles, $n=(\pm 1,0,0,0)$. At these points $\trace [U]$ and $\trace
[U^{2}]$ are invariant under first-order changes in the
eigenvalues. This would correspond to adding a small $\delta\vec{n}$
part to $\vec{n}$, but the change in $n_0$ is second order in this variable,
and the imaginary parts cancel to first order. Thus near the poles one
of the two coordinates is superfluous.

This becomes even clearer once we realise that the trace variables are
equivalent to the parametrisation through the use of diagonal
matrices, with the two parameters $n^{0}=(e^{i\psi _{1}}+e^{i\psi
_{2}})/2=e^{i\Psi}\cos(\psi _{1}-\psi _{2})/2$ and $n^{3}=i(e^{i\psi
_{1}}-e^{i\psi _{2}})/2=e^{i\Psi}\sin (\psi _{1}-\psi _{2})/2$.
Clearly all points with $\psi _{1}$ equal to $\psi _{2}$ correspond to
identical eigenvalues, and for those points one coordinate is
superfluous.  That the behaviour is like that of polar coordinates in
three dimensions is not obvious from this discussion, and can only be
found in an explicit calculation.

We have not yet discussed the integration measure, but it can be shown
(see Ref.~\cite{Weyl} and the discussion in Appendix \ref{app:xx})
that it takes the form we would expect from the kinetic energy
operator,
\begin{equation}
\left\langle \Psi _{1}\right| \left. \Psi _{2}\right\rangle =\frac{1}{(2\pi )^{N}}\int _{V}\Psi _{1}(\{\psi \})^{*}\Psi _{2}(\{\psi \})\, \Xi ^{2}\prod _{i}d\psi _{i}.
\end{equation}
Here we have not ordered the eigenvalues of $U$ in any way. Since an interchange
of the eigenvalues corresponds to the same class of matrices, the wave function
must be invariant under this interchange. Actually, it must be symmetric, and
the wave function is that of a system of $N$ bosons on a torus. However,
the complications of norm and kinetic energy suggest that it would be beneficial
to make the change of functions
\begin{equation}
\Phi (\{\psi \})\equiv \Xi \Psi (\{\psi \}),
\end{equation}
to eliminate the first derivatives from the kinetic energy, and the integration
weight from the norm. As usual this leads to interesting boundary conditions
on the function, since they must vanish where $\Xi $ vanishes, and since
$\Xi $ vanishes linearly and is antisymmetric under interchange of any
pair of $\psi $'s, the wave function $\Phi $ is antisymmetric in the
same way. This transformation thus turns the bosonic problem with a
complicated integration measure into a simpler fermionic problem,
\begin{eqnarray}
\sum _{a=1}^{N^{2}-1}E^{a}E^{a} & \rightarrow  & 
\Xi \left( \sum _{a=1}^{N^{2}-1}E^{a}E^{a}\right) \frac{1}{\Xi }=
D_{N}-\frac{N(N^{2}-1)}{24}\quad,
\nonumber\\
D_{N} & = & -\frac{1}{2}
\sum _{i}\frac{\partial^{2}}{\partial \psi _{i}^{2}}
\quad.
\label{Dn} 
\end{eqnarray}
Even though it would appear to be quite complicated to determine the
eigenfunctions of the quadratic Casimir operator (i.e., the square of
$E^{a}$), we show in Appendix \ref{app:xx} that we can label each
eigenfunction with a partition $[f]\equiv[f_{1},\ldots ,f_{N}]$
\begin{equation}
\Phi _{[f_{1},\ldots ,f_{N}]}=\left| \begin{array}{cccc}
e^{i(f_{1}+(N-1)/2)\psi _{1}} & e^{i(f_{2}+(N-3)/2)\psi _{1}} 
	& \cdots  & e^{i(f_{N}-(N-1)/2)\psi _{1}}\\
\vdots  & \vdots  & \ddots  & \vdots \\
e^{i(f_{1}+(N-1)/2)\psi _{N}} & e^{i(f_{2}+(N-3)/2)\psi _{N}} 
& \cdots  & e^{i(f_{N}-(N-1)/2)\psi _{N}}
\end{array}\right|\quad,
\end{equation}
where $[f_{1},\ldots ,f_{N}]$ is the standard $U(N)$ representation
labelled by a Young tableau, satisfying $f_{i}\geq f_{i+1}$.  Note
that we do not require that all $f$'s are positive or zero. More
details on the representation of the symmetric wave function as a
one-dimensional fermion problem can be found in Appendix \ref{app:xx},
where we prove that all excitations $\Phi _{[f]}$ can be represented in
terms of trace variables.

Even though this is all quite elegant, we are really only interested
in $SU(N)$, rather than $U(N)$. Fortunately it is very simple to show
that
\begin{equation}
\Phi _{[f]}=e^{i\sum _{j}f_{j}\Psi }\Phi _{[f]}(\{\psi _{i}-\Psi \})\quad,
\end{equation}
where
\begin{equation}
\Psi =\frac{1}{N}\sum _{j}\psi _{j}
\end{equation}
is the coordinate describing the phase of the determinant of $U$. If
we interpret $\Psi $ as a centre-of-mass coordinate, this shows that
the eigenfunctions of the kinetic energy (electric energy) separate
into a trivial centre-of-mass part, and a part that does not depend on
this coordinate at all, since it only depends on the distance between
the centre of mass and the individual particles.  This can also be
written, as we have shown explicitly in the case of $\Xi$, in terms
of the relative angles $(\psi _{i}-\psi _{j})$. This is clearly a simple
consequence of the separability of the kinetic energy in
centre-of-mass and relative coordinates.

The only well-known problem arising from this separability is that not
all functions $\Phi _{[f]}$ give rise to different functions of the
relative coordinates. Thus the solutions $\Phi _{[f]}$ where $[f]$ is
a partition of length $N-1$ form a complete set of polynomial
solutions, as a partition of length $N$ can be mapped onto partition
of shorter length since $\det U=1$, as we can see from the relation,
\begin{equation}
\Phi _{[f_{1}+n,\ldots \, f_{N}+n]}=e^{inN\Psi }\Phi _{[f_{1},\ldots, f_{N}]}
\quad ,
\label{det1}
\end{equation}
in which $n\in \mathbb {Z}$.

This knowledge is useful when constructing the full spectrum and all the eigenvalues
of the electric part of the Hamiltonian, 
\begin{equation}
\left( D_{N}-\frac{N(N^{2}-1)}{24}\right) \Phi _{[f]}=E^{0}_{[f]}\Phi _{[f]}\quad ,
\end{equation}
where the eigenvalues $E^{0}_{[f]}$ follow from Eq.~(\ref{Dn}),
\begin{eqnarray}
E^{0}_{[f]} & = & -\frac{N(N^{2}-1)}{24}+
\sum _{k=1}^{N-1}\frac{(f_{k}+(N-2k+1)/2)^{2}}{2}+
\frac{1}{8}(N-1)^{2}
-\frac{1}{2N}\left( \sum _{k=1}^{N-1}f_{k}\right) ^{2}\nonumber\\
 & = & \frac{N-1}{2N}\sum _{i=1}^{N-1}f_{k}^{2}+
\frac{1}{2}\sum _{k=1}^{N-1}(N-2k+1)f_{k}-
\frac{1}{N}\sum _{k<l}f_{k}f_{l}\quad .
\end{eqnarray}
Even though one can gain considerable insight from this separation, we
shall generally prefer to work with systems described in terms of the
``single-particle'' coordinates $\psi _{i}$, rather than the
centre-of-mass and relative coordinates, since it is very complicated
to deal with anti-symmetry in these latter coordinates. This is well known in
the translationally-invariant treatment of the finite many-body problem,
as can be found in standard nuclear physics text books (see, e.g., 
Ref.~\cite{RingSchuck}).

As seen above there are no problems when dealing with the electric
energy. It is very naturally expressed in terms of the momentum
operators $p_i=-\frac{\partial}{\partial \psi_i}$ as $H_E=\frac{1}{2}\sum_i
p_i^2$.  The picture changes once we look at the magnetic term in the
Hamiltonian which is easily expressed in terms of the eigenvalues of
$U$, the variables conjugate to $p_i$, as
\begin{eqnarray}
H_{M} & = & 2N-\trace [U]-\trace [U^{-1}]\nonumber\\
 & = & 2\sum _{i=1}^{N}(1-\cos \psi _{i})\quad.
\label{Hm} 
\end{eqnarray}
There seems to be no easy way to separate this part of the Hamiltonian in terms
of a relative and centre-of-mass part, and further investigation shows that
the commutator of $P\equiv\sum _{i}p_{i}$, the centre-of-mass momentum, with
the magnetic term is equally non-trivial,
\begin{equation}
\left[P,\sum _{i}\cos \psi _{i}\right]=i\sum _{i}\sin \psi _{i}\quad , 
\end{equation}
and thus the potential energy term also breaks translational
invariance.  The Hamiltonian does not commute with the constraint
$P=0$, which we need to impose on the $SU(N)$ problem, since $\Psi$ is
an unphysical variable. If we deal with this through the standard
technology of Dirac quantisation \cite{Dirac}, where we first
classically replace the Poisson bracket by the Dirac bracket
\begin{equation}
\{A,B\}_{\mathrm{DB}}=\{A,B\}-\{A,P\}\{\Psi,B\}+\{A,\Psi
\}\{P,B\}\quad,
\end{equation}
defined so that the Dirac brackets of the constraints $\Psi =0$ and
$P=0$ with the Hamiltonian are zero, and then quantise the theory by
replacing the Dirac bracket by a commutator, we find the relations
(we use a tilde to denote the variables obeying the Dirac bracket)
\begin{eqnarray}
[\tilde p_{i},\tilde p_{j}] & = & 0\quad,\nonumber\\
{}[\tilde \psi _{i},\tilde \psi _{j}] & = & 0\quad,\nonumber\\
{}[\tilde p_{i},\tilde \psi _{j}] & = & -i(\delta _{ij}-\frac{1}{N})\quad.
\end{eqnarray}
These relations can be satisfied by ordinary commutators if we relate the 
$SU(N)$ variables
$\tilde \psi$ to the  $\psi$ variables for $U(N)$ by
$\tilde \psi = \psi-\Psi$, equating $\tilde p_i$ to $p_i$. 
In that case the realisation of the Hamiltonian
\begin{equation}
H={2}\sum p_{i}^{2}+2 \lambda \sum _{i}[1-\cos (\psi _{i}-\Psi
)]\quad ,
\end{equation}
is explicitly translationally invariant, and we can separate all the
eigenstates into a $U(1)$ part and an $SU(N)$ part. An additional
factor 4 in the kinetic energy arise from the four separate links
around a plaquette. The Hamiltonian
is scaled with $g^2 a^{-1}$, such that it maps on an $N$-body
problem with unit masses  and  we introduce the coupling 
constant $\lambda = g^{-4}$, for $D=3$.

\subsection{\protect$\trace [U]\protect$ as a raising operator}

\begin{figure}
\centerline{\resizebox*{8cm}{!}{\includegraphics{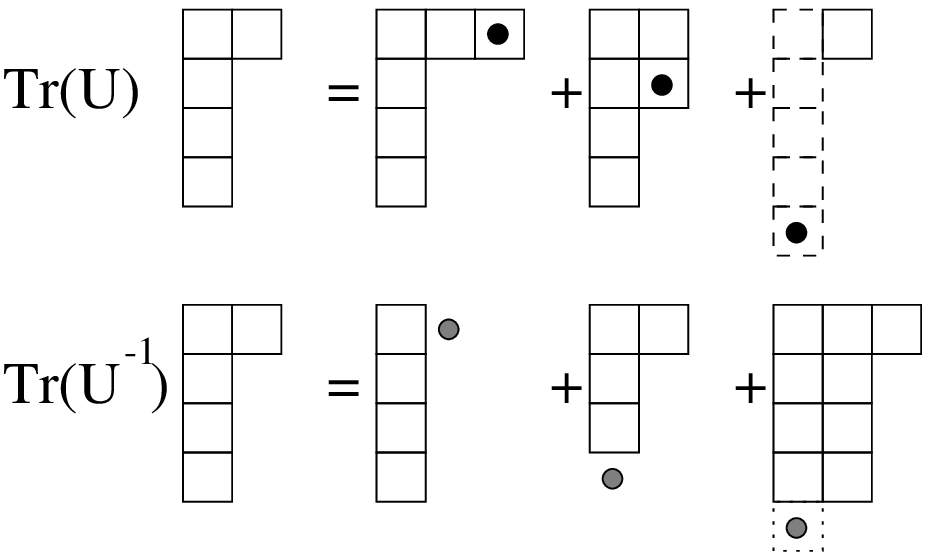}}}
\caption{An example of the multiplication of the state $[2,1^3]$ in
$SU(5)$ with \protect$\trace [U]\protect$ and \protect$\trace
[U^{-1}]\protect$. This corresponds to the addition or subtraction of
a square, such that a higher row is always larger than or equal to a
lower one. A full column can be added or subtracted without changing the
state. This is used to obtain a canonical form with 
the last column empty.}
\label{fig:young}
\end{figure}

We can of course work directly in terms of the angular variables as
discussed above. For some cases it is actually useful to consider
$\trace [U]$ and its complex conjugate as raising and lowering
operators, which act in a simple way on the Young tableau labelling
the symmetry of the wave function,
\begin{eqnarray}
\trace [U]\Phi _{[f]} & = & 
\sum _{i=0}^{N-1}\Phi _{[f_{1},\cdots ,f_{i}+1,
\cdots ,f_{N-1},f_{N}]}\quad ,\nonumber\\
\trace [U^{-1}]\Phi _{[f]} & = & 
\sum _{i=0}^{N-1}\Phi _{[f_{1},\cdots ,f_{i}-1,
\cdots ,f_{N-1},f_{N}]}\quad ,
\end{eqnarray}
where $f_{N}=0$ can be imposed, as discussed above. If at any stage in
the calculation we end up with $f_{N}\not =0$, we can use Eq.~(\ref{det1})
to make it zero. In summary, we need to use the  three conditions
\begin{eqnarray}
(f_{1},f_{2},\cdots ,f_{N-1},-1) & = & (f_{1}+1,f_{2}+1,\cdots
,f_{N-1}+1,0)\quad,
\nonumber\\
(n+f_{1},n+f_{2},\cdots ,n+f_{N-1},n) & = & 
(f_{1},f_{2},\cdots ,f_{N-1},0)\quad,\nonumber\\
(f_{1},f_{2},\cdots ,f_{i},\cdots ,f_{i},\cdots ,f_{N-1}) & = &
0\quad.
\end{eqnarray}
Even though this interpretation of the traces is quite appealing, it
is not totally straightforward to use, since there is no cyclic
vector; when acting on the state labelled by
\begin{equation}
[f]=[0^{N-1}]\quad,
\end{equation}
both terms lead to non-vanishing results. This is intimately related to the
existence of both covariant and contravariant representations of $SU(N)$
occurring at the same kinetic energy, which are mixed by the magnetic terms.
An example is shown in Fig.~\ref{fig:young}.

\subsection{The harmonic approximation}

In the weak-coupling limit we can make the harmonic approximation,
where we approximate the magnetic potential by its quadratic
expansion, and we can ignore the periodicity of the $\psi$ variables
since they oscillate near their equilibrium values. The resulting potential,
\begin{eqnarray}
V(\{\psi\}) &=&{\lambda} \sum_{i=1}^N (\psi_i-\Psi)^2\nonumber\\
            &=&{\lambda}\left(\sum_{i=1}^N \psi^2 -N\Psi^2\right)\quad,
\end{eqnarray}
is modified by confining the centre-of-mass in the harmonic oscillator
(HO) potential $V(\Psi)=\frac{\lambda}{2} N \Psi^2$. We have
reconstructed here the well-known separability of the many-body
harmonic oscillator Hamiltonian into a centre-of-mass (CM) and
relative part, where the CM part is in a harmonic oscillator state
with frequency $N$ times larger (since the mass for the CM mode is
also $N$ times the single-particle mass) than the single particle one
(and see, e.g., Ref.~\cite{Moshinski}).

Amongst the spectrum of the $N$-fermion HO problem we have states with
the CM in the $m=0,1,2,\ldots$ states, and we thus reproduce each relative
state with all of the CM states. If we can determine all those states
with the CM in its ground state, we immediately know the degeneracy of
the $SU(N)$ spectrum. So we need to identify those states from amongst
the degenerate multiplet with energy $(n+N/2)\hbar\omega$. Let $\nu_n$
be the degeneracy of this ($n$-th excited) energy level, which is
given by the number of distinct partitions of $n$ into $N$ positive
numbers:
\begin{equation}
\nu_n^{U(N)} =P_N(n) =
\sum_{\lambda_1>\lambda_2>\cdots>\lambda_N\geq 0; \sum \lambda_i=n}1 \quad.
\end{equation}
We use the fact that harmonic oscillator wave functions are
exponentials times a polynomial. Clearly the centre-of-mass polynomial
can only arise from a linear combination of the polynomial parts of
the various degenerate wave functions. For a CM ground state this
polynomial part is a constant, and thus lies in the null space of the
centre-of-mass momentum operator
\begin{equation}
P = \sum_{i=1}^N i \frac{\partial}{\partial \psi_i}  \quad.
\end{equation}
This operator maps the $n$-th degree polynomials onto the $(n-1)$-th
degree polynomials. The dimension of the null space is the difference
between the dimensions of the domain and range of the operator $P$.
Therefore the degeneracy of the $n$-th relative motion, i.e.,
$SU(N)$, eigenstate is given by
\begin{equation}
\nu_n^{SU(N)} = P_N(n)-P_N(n-1)  \quad.
\end{equation}
For $N=2$
\begin{equation}
P_2(n)  =  \left[\frac{n+1}{2} \right]  \quad, 
\end{equation}
where $[\ \cdot\ ]$ denotes the integer part of a number, and thus
$\nu_n = \delta_{n,\text{odd}}$, which corresponds to 
a set of non-degenerate, odd wave functions for $SU(2)$.
For $SU(3)$, $SU(4)$, and $SU(5)$ we find the degeneracies (the first
state always occurs at $i=N(N-1)/2$, since all lower energy functions
are not antisymmetric),
\begin{eqnarray}
\{P_3(i)\}_{i=3}^\infty & = & 
\{ 1, 1, 2, 3, 4, 5, 7, 8, 10, 12, 14, \cdots \}  \quad, \nonumber\\
\{\nu_i^{SU(3)}\}_{i=3}^\infty & = & 
\{ 1, 0, 1, 1, 1, 1, 2, 1, 2, 2, 2, \cdots \}  \quad, \nonumber\\
\{P_4(i)\}_{i=6}^\infty  & = &
\{ 1, 1, 2, 3, 5, 6, 9, 11, 15, 18, \cdots \}\quad, \nonumber\\
\{\nu_i^{SU(4)}\}_{i=6}^\infty & = & 
\{ 1, 0, 1, 1, 2, 1, 3, 2, 4, 3 \cdots \}  \quad, \nonumber\\
\{P_5(i)\}_{i=10}^\infty  & = &
\{1, 1, 2, 3, 5, 7, 10, 13, 18, 23, 30, \cdots \} \quad, \nonumber\\
\{\nu_i^{SU(5)}\}_{i=10}^\infty & = & 
\{1, 0, 1, 1, 2, 2, 3, 3, 5, 5, 7, \cdots \}  
\quad. \label{eq:hanu}
\end{eqnarray}
Notice the absence of the first excited state; $\nu_{N(N-1)/2+1}^{SU(N)} =0$,
since the only state at that position is a centre-of-mass excitation
built on the ground state.
Note that our result differs from the result in
Ref.~\cite{RW80}, where the fact
that only even functions in the radial coordinate are allowed was ignored.
The $SU(3)$ spectrum is
\begin{equation}
\frac{2a}{g^2} E_x \sim 2 \sqrt{2} \frac{1}{g^2} (3k + 2n) \quad,
\end{equation}
for $n=0,1,2,3, \ldots$ and $k=1,2,3,\ldots$.

\subsection{Large-$N$ limit}
The large-$N$ limit of QCD is one of the intriguing open problems in
the theory, and one might wonder whether our result can shed some
light on the much less taxing problem of the large-$N$ limit of the
one-plaquette problem. In that case we have to study the Hamiltonian 
\begin{equation}
H= {2} \sum_{i} p_i^2 + 2\lambda\sum _{i=1}^{N}(1-\cos (\psi_i-\Psi)
)\quad.
\end{equation}
The leading term is simply obtained from the Hartree-Fock
approximation, with Hartree Hamiltonian
\begin{equation}
\mathcal{H}={2} p^2+2\lambda(1-\cos \psi)\quad,
\end{equation}
and we can (if we wish) evaluate the Hartree-Fock energy for this
state using the solutions of the Mathieu equation. The boundary
conditions are strict periodicity, since the all the circles
parametrised by $\psi_i$ are
connected, and end up at the same physical point.

Such results are well-known from many-body theory \cite{BlaizotRipka}.
In principle the next order in $1/N$ would follow from an RPA
calculation. However, there are some problems. The interaction is
explicitly $N$-dependent, since $\Psi$ depends on $N$, and the
residual interaction is rather cumbersome to deal with.

\subsection{Explicit solutions}
\begin{figure}
\centerline{\resizebox*{6cm}{!}{\includegraphics{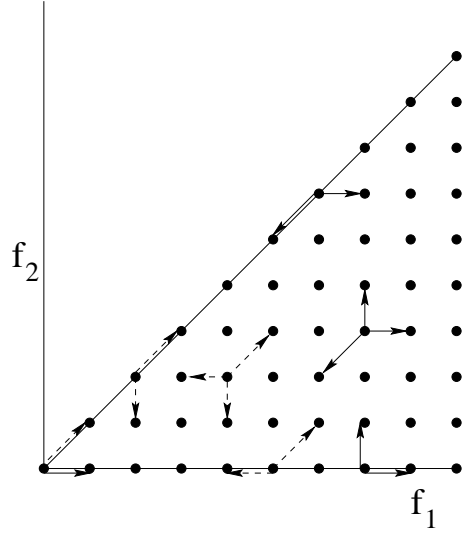}} }
\caption{Examples of the action of $\trace U$ (solid lines with arrows) 
and $\trace[U^\dagger]$ (dashed lines with arrows) on the states
of $SU(3)$, labelled by the partition $[f_1,f_2]$.}
\label{fig:su3step}
\end{figure}
Let us now investigate the spectra of a few of the relevant gauge
theories. The method we use to solve the problem is to work in a basis
of eigenstates of the electric Hamiltonian, and evaluate the action of
$\trace U$ and $\trace U^\dagger$ on these states. As can be seen from
the example for $SU(3)$ shown in Fig.~\ref{fig:su3step}, there are
several different coupling mechanisms. Nonetheless, this method is
straightforward to implement, and leads to fully converged results up
to large values of the effective coupling constant $1/g^4$.

\subsubsection{The $U(1)$ one-plaquette problem}

\begin{figure}
\centerline{\resizebox*{10cm}{!}{\includegraphics{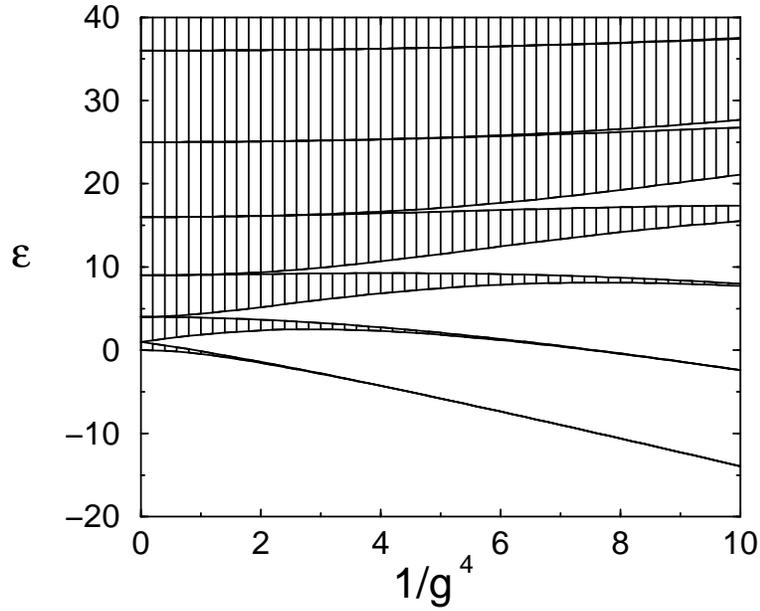}} }
\caption{The lower end of the fully converged $U(1)$ one-plaquette
spectrum.  The lines starting from $4 n^2$ are the periodic solutions, and the
lines starting from $(2 n+1)^2$ are anti-periodic ones.
The hashed areas denote the existence
of a continuum of Floquet states at these energies.} \label{fig:u1}
\end{figure}

First we look at $U(1)$.  This is something of a special case, as it does
not fit into our general $SU(N)$ framework. However, the relation is
straightforward.  With the parametrisation $U=e^{i\phi}$, the electric
operator is $i \partial /\partial \phi $. If we use the maximal tree
formulation, we find that the action of the electric energy gets
weighted by a factor of four, essentially one for each side of the
plaquette. The magnetic part is $[2-\exp(i\phi) -\exp (-i\phi)]
=2(1-\cos \phi )$, and hence the Schr\"{o}dinger equation becomes
\begin{equation}
\frac{g^{2}}{2a}\left[ -4\frac{d ^{2}}{d \phi ^{2}}+
\frac{2}{g^{4}}(1-\cos \phi )\right] \Psi =E\Psi \quad.
\end{equation}
Upon substitution of $2z=\phi$, $\alpha=2(Ea-1)/g^2$, $q=-2/g^2$, we
find the Mathieu equation,
\begin{equation}
\frac{d ^{2}y(z)}{d z^{2}}+(\alpha-2q\cos 2z)y(z)=0\quad.
\end{equation}
However, we still need to analyse the boundary conditions on
$\Psi$. It is well known from Floquet theory \cite{WW27} that solutions of a
differential equation with periodic coefficients are quasi-periodic,
\begin{equation}
\Psi _{\nu }=e^{i\pi \nu }\Psi \quad,
\end{equation}
where $\Psi (\phi )=\Psi (\phi +\pi )$, and $\nu$ is a, in principle
complex, exponent.  The only normalisable solutions are those with a
real exponent, and those are found bracketed between the (periodic)
$\ce _{r}(z,q)$ and the $\se _{r+1}(z,q)$ solutions \cite{AS64}, as in
Fig.~\ref{fig:u1}. These results are analogous to the band structure in
metals, due to the periodic potential in which the electrons move.
We define a scaled energy $\varepsilon$ as 
\begin{equation}
E = \frac{g^2}{2 a} \varepsilon + \frac{N}{g^2 a} \quad ,
\end{equation} 
which we use for representing the numerical results.

For the period states 
we can attack the problem numerically by expanding the Hamiltonian
in terms of the eigenstates, $\exp(in\phi)$, of the electric operator,
\begin{eqnarray}
H_{\ce } & = & \sum _{n=0}|n\rangle \left( 4 \frac{n^{2}g^{2}}{2a}+\frac{1}{g^{2}a}\right) \langle n|\quad-\quad\frac{1}{2g^{2}a}\big \{|n\rangle \langle n+1|\quad+\quad|n+1\rangle (1+\delta _{0n})\langle n|\big \}\\
H_{\se } & = & \sum _{n=1}|n\rangle \left( 4 \frac{n^{2}g^{2}}{2a}+\frac{1}{g^{2}a}\right) \langle n|\quad-\quad\frac{1}{2g^{2}a}\big \{|n\rangle \langle n+1|\quad+\, \, |n+1\rangle \langle n|\big \}
\end{eqnarray}
 where the sum starts at zero for $\ce _{2r}$ solutions and at one for
$\se _{2r}$ solutions. This matrix equation serves as a relatively efficient
way of determining the characteristic values $a$ and $b$ of the Mathieu
equation, via straightforward numerical methods.

\subsubsection{The $SU(2)$ one-plaquette problem}
\begin{figure}
\centerline{\resizebox*{10cm}{!}{\includegraphics{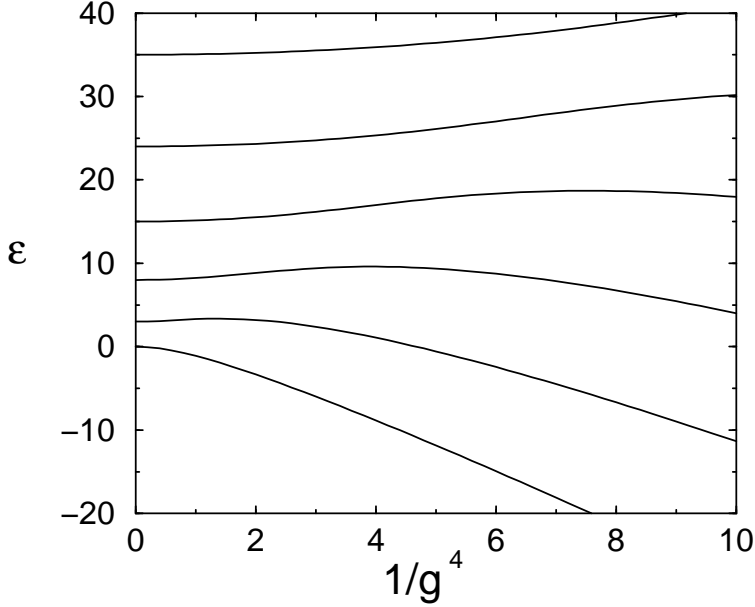}} }
\caption{The lower end of the fully converged $SU(2)$ one-plaquette spectrum}
\label{fig:su2}
\end{figure}

The $SU(2)$ matrices have two complex conjugate eigenvalues, and can
be parametrised as $\trace(U)=2\cos\phi$, where $\phi=(\psi_1-\psi_2)$, 
and see Eq.~(\ref{eq:Ueigval}). If we absorb the factor $\Xi$ from
Eq.~(\ref{eq:defXi}) into the wave function ($\Xi=2\sin(\phi/2)$),
\begin{equation}
\Phi=\Xi \Psi
\end{equation}
The  Schr\"{o}dinger equation, which takes the form
\begin{equation}
\frac{g^{2}}{2a}\left[ -4\frac{\partial ^{2}}{\partial \phi
^{2}}-1+\frac{4}{g^{4}}(1-\cos (\phi/2) )\right]  \Phi =E  \Phi \quad,
\end{equation}
is again the Mathieu equation. In this case the interpretation is
different, even though we still have $2z=\phi$, but $\phi$ plays a different
role.  As a  function of the periodic quantity $\trace U$, $\Phi$
must be strictly periodic. Due to the anti-symmetry induced by the Pauli
principle, the only non-singular wave functions $\Psi $ are the odd
Mathieu functions $\se _{1}(z,q),\se _{2}(z,q),\cdots $, for which
$\alpha=b_{r}(q)$, where the index is allowed to take odd values,
since the $2\pi $ periodicity on $[-\pi ,\pi ]$ is guaranteed by the
solutions being odd. For $q=0$ the solutions reduce to 
$\Phi _{n}=\sin \frac{n+1}{2}\phi $, and since $\Xi\propto \sin \frac{1}{2} \phi $,
 $\Psi (-\pi)=\Psi (\pi )$. The solutions are shown in Fig.~\ref{fig:su2}.

In terms of the eigenstates $|n\rangle $ of the electric operator the
Hamiltonian reads
\begin{equation}
H=\sum _{n=0}|n\rangle \left( \frac{n(n+2)g^{2}}{2a}+\frac{2}{g^{2}a}\right) \langle n|\quad-\quad\frac{1}{g^{2}a}\big \{|n\rangle \langle n+1|\quad+\quad|n+1\rangle \langle n|\big \}\quad.
\end{equation}
We see that the differences between the $U(1)$ case and the $SU(2)$
case lie in the diagonal part of the Hamiltonian, respectively $n^{2}$
and $(n+1)^{2}-1$, which result from the electric operator.

\subsubsection{$SU(3)$ one-plaquette problem}

\begin{figure}
\centerline{\resizebox*{8cm}{!}{\includegraphics{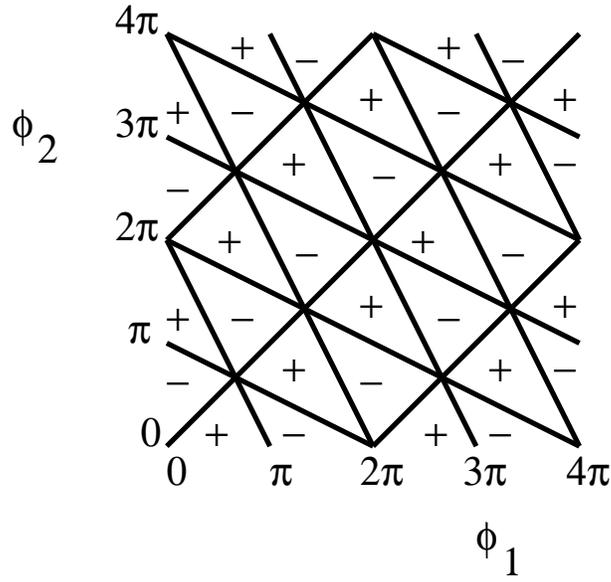}} }
\caption{The domains for \protect$\phi _{1}\protect$ and \protect$\phi
_{2}\protect$, in the antisymmetric wavefunction \protect$\Phi
\protect$. }
\label{fig:FundSU3}
\end{figure}

The $SU(3)$ one-plaquette problem has two independent angular degrees
of freedom.  From the $U(3)$ parametrisation we see that these can be
chosen as $\phi_1=\frac{1}{3}(2\psi_1-\psi_2-\psi_3)$, 
$\phi_2=\frac{1}{3}(-\psi_1+2\psi_2-\psi_3)$. The remaining
combination,
$\frac{1}{3}(-\psi_1-\psi_2+2\psi_3)=-(\phi_1+\phi_2)$, is linearly
dependent.

The Hamiltonian in these angular coordinates
becomes, using Eqs.~(\ref{Dn}) and (\ref{Hm}), 
\begin{equation}
\frac{g^{2}}{2a}\left[ \frac{4}{3}\left\{ -\frac{\partial
^{2}}{\partial \phi _{1}^{2}}-\frac{\partial ^{2}}{\partial \phi
_{2}^{2}}+\frac{\partial ^{2}}{\partial \phi _{1}\partial \phi
_{2}}-3\right\} +\frac{1}{g^{4}}(3-\cos \phi _{1}-\cos \phi _{2}-\cos
(\phi _{1}+\phi _{2}))\right] \Phi =E \Phi \quad,
\end{equation}
where the boundary conditions are determined from the anti-symmetry of
$\Psi $.  The wave function vanishes if two eigenvalues of the matrix
$U$ coincide, $\phi _{1}\not =\phi _{2}\not =-(\phi _{1}+\phi
_{2})\not =\phi _{1}$.  All variables are defined modulo $2\pi $, as
they correspond to identical eigenvalues. We can restrict ourselves to
the domain $-\phi _{1}-\phi _{2}\leq \phi _{1}$ , $\phi _{1}\leq \phi
_{2}$, and $\phi _{2}\leq 2\pi -\phi _{1}-\phi _{2}$.  However, many
other choices of boundary conditions are possible, which follow from
either permuting the variables, or translating the domain by $2\pi $.
The $\phi _{1}\otimes \phi _{2}$ plane is divided into copies of the
fundamental domain by the lines $\phi _{1}-\phi _{2}=n_{1}\pi $, $\phi
_{1}+\frac{1}{2}\phi _{2}=n_{2}\pi $, and $\frac{1}{2}\phi _{1}+\phi
_{2}=n_{3}\pi $, where $n_{i}\in \cdots ,-2,-1,0,1,2,\cdots $, as
shown in Fig.~\ref{fig:FundSU3}.

\begin{figure}
\centerline{\resizebox*{10cm}{!}{\includegraphics{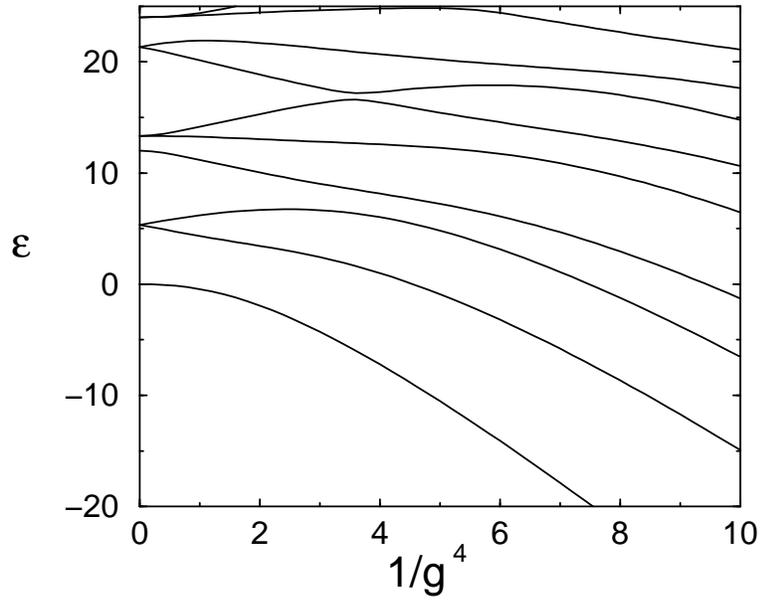}} }
\caption{The lower end of the fully converged $SU(3)$ one-plaquette spectrum}
\label{fig:specSU3}
\end{figure}

The potential can be simplified to
\begin{equation}
1+\cos \phi _{1}+\cos \phi _{2}+\cos (\phi _{1}+\phi _{2})=4\cos \frac{\phi _{1}}{2}\cos \frac{\phi _{2}}{2}\cos \frac{\phi _{1}+\phi _{2}}{2}\quad.
\end{equation}
The explicit angular Schr\"{o}dinger equation is useful for the harmonic approximation
in the weak-coupling limit, however, for the strong-coupling limit and the intermediate
region in coupling constants, the Hamiltonian in terms of the electric-operator
eigenstates $|f_1,f_2\rangle $ is more useful 
\begin{eqnarray}
H & = & \sum _{0\leq f_2\leq f_1}|f_1,f_2\rangle \left
( \frac{4 (f_1^{2}+f_2^{2}+3f_1-f_1 f_2)g^{2}}{6a}+
\frac{3}{g^{2}a}\right) \langle f_1,f_2|\nonumber\\
 &  & -\frac{1}{2g^{2}a}\big \{|f_1,f_2\rangle \langle f_1,f_2+1|\quad +\quad |f_1,f_2\rangle \langle f_1+1,f_2|\nonumber\\
 &  & +|f_1+1,f_2\rangle \langle f_1,f_2|\quad +\quad |f_1,f_2+1\rangle \langle f_1,f_2|\nonumber\\
 &  & +|f_1,f_2\rangle \langle f_1+1,f_2+1|\quad +\quad |f_1+1,f_2+1\rangle \langle f_1,f_2|\big \}\quad ,
\end{eqnarray}
where all the states $|f_1,f_2\rangle $ where the condition $0\leq f_2\leq f_1$ does not hold, are identical to zero. This
problem does not seem to have a closed-form analytical solution, and we have to
resort to numerical methods to obtain the spectrum for arbitrary values of the
coupling constant $g$, as in Fig.~\ref{fig:specSU3}.

\subsubsection{Higher orders}
\begin{figure}
\centerline{\resizebox*{10cm}{!}{\includegraphics{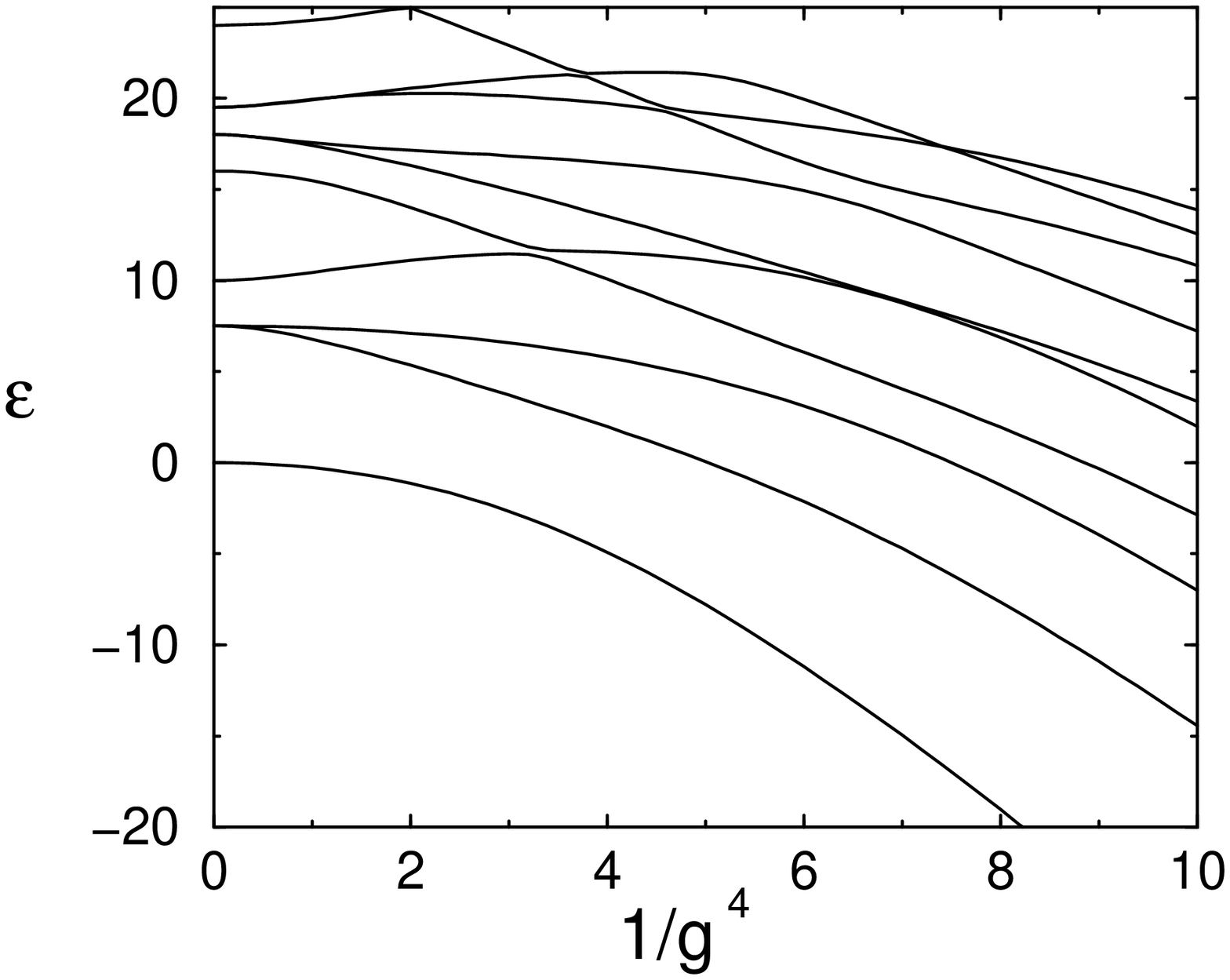}} }
\caption{The lower end of the fully converged $SU(4)$ one-plaquette spectrum}
\label{fig:su4}
\end{figure}

\begin{figure}
\centerline{\resizebox*{10cm}{!}{\includegraphics{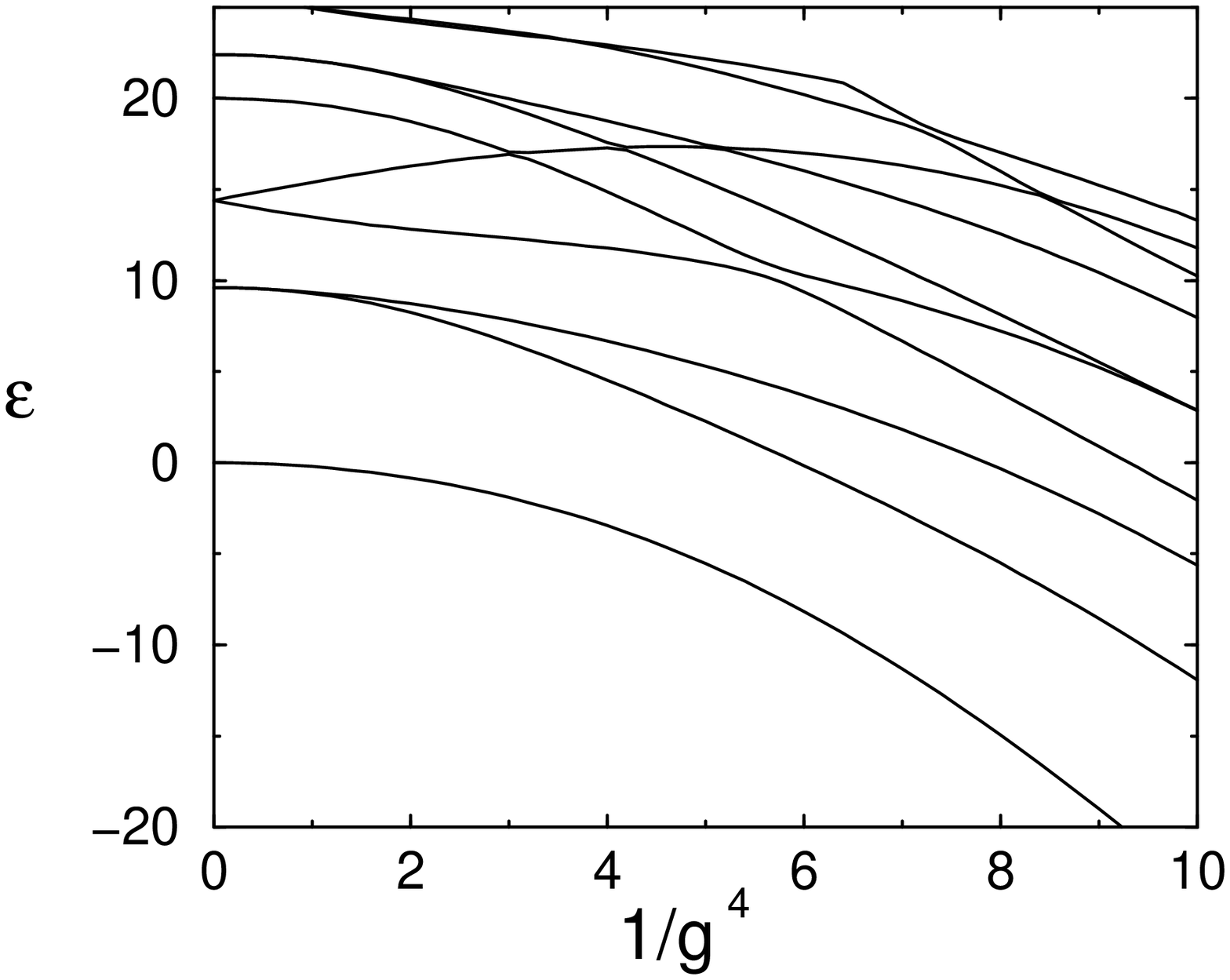}} }
\caption{The lower end of the fully converged $SU(5)$ one-plaquette spectrum}
\label{fig:su5}
\end{figure}

We conclude this section by showing results for the gauge groups
$SU(4)$ (Fig.~\ref{fig:su4}) and $SU(5)$ (Fig.~\ref{fig:su5}). The
most interesting aspect of these results is the presence of (what
appear to be) real level crossings, which could be a reflection of some
hidden symmetry in the problem. It was verified that the distance at
these crossings was  equal to zero within the numerical accuracy.

The region of the coupling constant shown in
Figs.~\ref{fig:u1}--\ref{fig:su5} is insufficient to see convergence
to the harmonic approximation (\ref{eq:hanu}). We have checked
numerically, for $g^{-4} \geq 50$, that these results are correct.

\section{Wave functionals}

The results for the one-plaquette problem have more consequences for
variational wave functionals than one might suspect. If the trace of
the one-plaquette matrix is used, the wave functional is a function of
the group characters only. The wave functional, which is the sum of
of one-plaquette functions,
\begin{equation}
\langle \{\psi _{\alpha i}\}|\Sigma \rangle =\sum _{\mathrm{plaquettes}\, \, \alpha }F\left( \{\psi _{\alpha i}\}_{i=1}^{N-1}\right) \quad\, \, ,
\end{equation}
 naturally leads to the sum of one-plaquette problems, leading to the energies
which are the sum one-plaquette energies. However, the product wave functional
\begin{equation}
\langle \{\psi _{\alpha i}\}|\Pi \rangle =\prod _{\mathrm{plaquettes}\, \, \alpha }F\left( \{\psi _{\alpha i}\}_{i=1}^{N-1}\right) \quad\, \, ,
\end{equation}
 also leads to the same result. Note that the exponentiated
potential, $\exp \{\kappa H_{M}\}$, where $\kappa $ is a parameter to
be determined by a variational calculation, as a simple ansatz for the
correlated wave functional \cite{Gre87} is a particular choice in this class of
wave functionals. The absence of correlations between
nearest-neighbour plaquettes, generally present, follows from the
symmetry in the angular variables,
\begin{equation}
\Psi _{\lambda }(\cdots \psi _{i}\cdots \psi _{j}\cdots )=\Psi _{\lambda }(\cdots \psi _{j}\cdots \psi _{i}\cdots )\quad,
\end{equation}
 where 
\begin{equation}
F(\psi _{1}\cdots \psi _{N})=\sum _{\lambda }c_\lambda 
\Psi _{\lambda }(\psi _{1}\cdots \psi _{N})\quad.
\end{equation}
 Therefore the product term from the electric operator vanishes, 
\begin{eqnarray}
D_{N}\Psi _{\lambda }'(\psi _{\alpha 1}\cdots \psi _{\alpha N})\Psi
 _{\lambda' }'(\psi _{\beta 1}\cdots \psi _{\beta N}) & = & \Psi
 _{\lambda' }'(\psi _{\beta 1}\cdots \psi _{\beta N})D_{N}\Psi
 _{\lambda }'(\psi _{\alpha 1}\cdots \psi _{\alpha N})\nonumber\\ & +
 & \Psi _{\lambda }'(\psi _{\alpha 1}\cdots \psi _{\alpha N})D_{N}\Psi
 _{\lambda' }'(\psi _{\beta 1}\cdots \psi _{\beta N})\quad\, \, ,
\end{eqnarray}
 where $\alpha $ and $\beta $ are plaquettes containing the link $l$,
and $\Psi _{\lambda }'=J\Psi _{\lambda }$. The differential operator
$D_{N}$ contains both sets of angular operators $\partial _{\psi
_{i}}=\partial _{\psi _{\alpha i}}+\partial _{\psi _{\beta i}}$.
Therefore the Hamiltonian, acting on the product wave functional $|\Pi
\rangle $ also reduces to the sum of one-plaquette Hamiltonians.

\section{Conclusions\label{sec:concl}}


With the understanding of both a gauge fixed Hamiltonian, and the
one-plaquette problem we would now like to use this in a method
designed to determine the ground-state wave function and excitations
from the Hamiltonian. Since there are, in principle, an infinite
number of degrees of freedom, some approximation is required. However,
we want to do this in a systematic manner, where we can improve the
order of approximation as required. The coupled cluster method (CCM)
\cite{CCMpapers}, a many-body technique, is suitable for such
calculations. It has previously been applied lattice field theories,
\cite{LWB}, and in various guises to gauge theory \cite{CCMFT},
although in a different context. It has also been applied to various spin-lattice problems in magnetism \cite{CCMmagnetism}.

Central to the CCM is the parametrisation of the ket-state wave
function as an exponential of excitation operators
\begin{equation}
|\Psi \rangle =e^{S}|\Phi \rangle \quad.
\end{equation}
The bra-state contains the inverse exponential such that the Hamiltonian functional
only contains linked terms of the excitation operators, due to the implicit
similarity transform $e^{-S}He^{S}$. The projection on the model state
$|\Phi \rangle $ is due to the annihilation operators $\tilde{S}$, 
\begin{eqnarray}
\langle \Psi | & = & \langle \Phi |(1+\tilde{S})e^{-S}\quad,\nonumber\\
\langle \Psi | & = & \langle \Phi |e^{\tilde{S}}e^{-S}\quad,
\end{eqnarray}
where the first form is used in the normal coupled cluster method
(NCCM), and the second form in the extended coupled cluster method
(ECCM). The normal coupled cluster method does not reproduce the
correct scaling limit when the excitation operators are obtained from
the strong-coupling expansion.

However, this and the extreme difficulty of implementing the ECCM are
related to the functional implementation of the CCM, where an operator
is simply chosen to be the multiplication of the wave function with a
function. We will only consider the operatorial CCM, where the
creation operators excite from the model state to any arbitrary
state, and do not act between different excited states on overlapping
lattice-site configurations
\begin{equation}
S=\sum _{I}c_{I}|I\rangle \langle \Phi |\quad.
\end{equation}
 This allows us to implement the ECCM, which is better at describing a system
at large global changes away from the model state $\Phi $.

The central question is the form of the states $|I\rangle $ in which
to expand the Hilbert space. This is the question discussed in the
first part of this paper.  For the colourless sector we have to use
closed contours, which are traces over products of $X$ variables,
since only these variables are invariant under gauge transformations
generated by Gauss' law. The traces over $X$ variables are
automatically closed contours as each contour associated with an $X$
variable starts and ends at the origin.

Our Hamiltonian is different from the naive Kogut-Susskind Hamiltonian
used in Refs.~\cite{Schuette,Leo98,GL97}, and this should have an
important effect on the results we hope to obtain in our future
work. The price we have to pay is that our Hamiltonian has a preferred
direction, and we have lost explicit translational invariance. It is
probably worthwhile to investigate different, more symmetrical,
choices of the maximal tree. It would seem that the current choice
should be very good to consider the interaction between fixed sources,
which clearly break translational symmetry, but it is less obviously
optimal for a study of the vacuum.

\section*{Acknowledgements}
This work was supported by a research grant (GR/L22331) from
the Engineering and Physical Sciences Research Council (EPSRC)
of Great Britain.\\
We thank Prof. Michael Thies for a useful discussion 
of the fermionic treatment of the one-plaquette problem.

\appendix

\section{One-dimensional many-fermion systems and their representations}
\label{app:xx}

As the ${SU}(N)$ one-plaquette problem can be mapped onto a one-dimensional
$(N-1)$-fermion problem, we analyse the fermion problem in some depth to
determine the spectrum and the action of a potential in coordinate representation.
Usually fermion problems are dealt with in operator form, here we use the knowledge
of (anti)symmetric functions to formulate the problem.

An $n$-fermion wave function is in terms of single-particle orbitals
$\phi_i(x)$ given by the fully anti-symmetrised Slater determinant
\begin{equation}
\langle x_1,\ldots,x_N | \Psi\rangle =
\begin{vmatrix}
\phi _{1}(x_{1})&\phi _{2}(x_{1})&\phi _{3}(x_{1})&\cdots& \phi _{n}(x_{1})\\
\phi _{1}(x_{2})&\phi _{2}(x_{2})&\phi _{3}(x_{2})&\cdots& \phi _{n}(x_{2})\\
\vdots &\vdots &\vdots &\cdots &\vdots\\
\phi _{1}(x_{n})&\phi _{2}(x_{n})&\phi _{3}(x_{n})&\cdots &\phi _{n}(x_{n})
\end{vmatrix}\quad.
\end{equation}
If $\phi _{m}(x)\equiv w(x)P_{m}(x)$ are chosen to be the lowest set
of single-particle orbitals, in which $w$ is the weight function,
which is the square root of the integration measure, and $P_{m}$ is
the $m$-th orthogonal polynomial given that weight, then the wave
function $\Psi$ is the ground state wave function $\Phi_0$. Only the
linearly independent parts of each column in the Slater determinant
contribute, and these are just the highest powers in each polynomial
$P_{m}(x)\rightarrow x^{m}$.  Therefore the ground-state Slater
determinant reduces to the Vandermonde determinant
\begin{equation}
\langle x|\Phi _{0}\rangle =\frac{1}{\sqrt{n!}}\prod _{i=1}^{n}w(x_{i})\prod _{k<l}^{n}(x_{k}-x_{l})\quad.
\end{equation}

As examples we deal with the infinite square well $[-\pi /2,\pi /2]$
and the harmonic oscillator. The one-particle wave functions are given,
respectively, by
\begin{eqnarray}
\{\phi \}_{\mathrm{s}.\mathrm{w}.} & = & \frac{2}{\sqrt{\pi }}\sin 2nx;\quad\frac{2}{\sqrt{\pi }}\cos (2n+1)x\quad,\\
\{\phi \}_{\mathrm{h}.\mathrm{o}.} & = & \frac{1}{\sqrt{2^{n}n!\sqrt{\pi }}}\eulere ^{-x^{2}/2}H_{n}(x)\quad.
\end{eqnarray}
 The sines and cosines can be re-expressed as polynomials in $\sin x$ with
the weight $\cos x$. The infinite domain of the harmonic oscillator is
of no consequence as the weight damps the wave function at large distances.
The non-interacting $n$-fermion ground state consists of fermions in 
the lowest $n$
levels. We find 
\begin{eqnarray}
\langle x|\Phi _{0}\rangle _{\mathrm{s}.\mathrm{w}.} & = & \left( \frac{2}{\pi }\right) ^{n/2}\frac{2^{[n/2]}[n/2]!}{\sqrt{n!}}\prod _{i=1}^{n}\cos x_{i}\prod _{k<l}^{n}(\sin x_{k}-\sin x_{l})\quad,\\
\langle x|\Phi _{0}\rangle _{\mathrm{h}.\mathrm{o}.} & = & \frac{1}{\pi ^{n/4}}\left( \prod _{j=1}^{n-1}\sqrt{2^{j}j!}\right) ^{-1}\frac{1}{\sqrt{n!}}\prod _{i=1}^{n}\eulere ^{-x_{i}^{2}/2}\prod _{k<l}2(x_{k}-x_{l})\quad,
\end{eqnarray}
 where the normalisation of the square-well wave function follows from De Moivre's
formula, and the normalisation of the harmonic oscillator wave function is a
direct consequence of the normalisation of the Hermite polynomials $H_{n}$.

We now consider an arbitrary excited state $\Phi _{\lambda }$. Since
the anti-symmetry is present by the ground-state wave function, the
wave function of an excited state can be factorised as a symmetric
function of all the variables times the ground-state wave
function. For any finite basis, this symmetric function must be a
symmetric polynomial
\begin{equation}
\langle x|\Phi _{\lambda }\rangle =S_{\lambda}(x)\langle x|\Phi _{0}\rangle \quad.
\end{equation}
Now the possible excitations follow from the theory of symmetric
polynomials, and can be labelled as a distinct partition $\lambda
=(\lambda _{1},\lambda _{2}\, \cdots ,\lambda _{n})$, where $\lambda
_{i}$ is the label of an occupied level, in ascending order.

\begin{figure}
\centerline{\resizebox*{10cm}{!}{\includegraphics{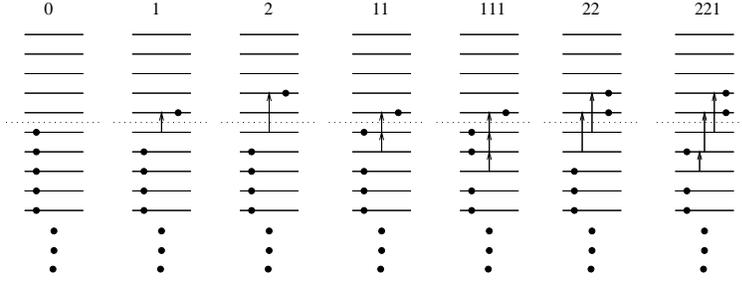}} }
\caption{The labelling of excitations in a fermionic system as non-distinct 
partitions,
where any partition \protect$\lambda _{1}\lambda _{2}\cdots \protect$ for
which \protect$\lambda _{i}\geq \lambda _{j}\protect$ if \protect$i<j\protect$
represents a unique excited state \protect$|n(n-1)(n-2)\cdots \rangle 
\rightarrow |(n+\lambda _{1})(n-1+\lambda _{2})
(n-2+\lambda _{3})\cdots \rangle \protect$,
where the labelling of states is the distinct partition. }
\label{fermiex}
\end{figure}

The symmetric polynomials are usually labelled independently of the
degree of filling $n$ by the non-distinct partition $f =\lambda
-\delta $ where $\delta =(n,n-1,\cdots ,2,1)$ such that 
$f _{i}\leq f _{j}$ for $i>j$. The non-distinct partition $f$
yields the excitation with respect to the Fermi level (see
Fig.~\ref{fermiex}).

Generally, the symmetric functions $S_{\lambda }$ are not of a definite
degree and they have a normalisation that depends on $n$. So $\lambda $
refers to the leading order polynomial $S_{\lambda }=cs_{\lambda }+\cdots $.
These polynomials $s_{\lambda }$ are the Schur functions \cite{McD79},
\begin{eqnarray}
s_{1} & = & \sum _{i}x_{i}\quad,\nonumber\\
s_{2} & = & \sum _{i}x_{i}^{2}+\sum _{i<j}x_{i}x_{j}\quad,\nonumber\\
s_{1,1} & = & \sum _{i<j}x_{i}x_{j}\nonumber\\
s_{3} & = & \sum _{i}x_{i}^{3}+\sum _{i<j}(x_{i}^{2}x_{j}+x_{j}^{2}x_{i})+\sum _{i<j<k}x_{i}x_{j}x_{k}\quad,\nonumber\\
s_{2,1} & = & \sum _{i<j}(x_{i}^{2}x_{j}+x_{j}^{2}x_{i})+2\sum _{i<j<k}x_{i}x_{j}x_{k}\quad,\nonumber\\
s_{1,1,1} & = & \sum _{i<j<k}x_{i}x_{j}x_{k}\quad.
\end{eqnarray}

For ${SU}(N)$ one-plaquette problem the excited states are symmetric
polynomials in the eigenvalues of the matrix $U\in {SU}(N)$. These
polynomials are the Schur functions, $S_{\lambda }=s_{\lambda }$, and,
as the eigenvalues lie on the complex unit circle, the symmetric
polynomials are properly normalised. The eigenvalues of the electric
operator follow directly from the definition. The two remaining
problems are the action of the magnetic term on these states and how,
and if, to express the symmetric polynomials, in the eigenvalues,
again in terms of trace variables. For this purpose we introduce three
representations: the Schur functions $s$, the monomial symmetric
functions $m$, and the so-called trace functions $t$,
\begin{eqnarray}
s_{\lambda } & = & \frac{\det x_{j}^{i-1+\lambda _{N-i+1}}}{\det x_{j}^{i-1}}\quad,\nonumber\\
m_{\lambda } & = & \sum _{\mathrm{Perm}(i_{j})}x^{\lambda _{1}}_{i_{1}}x^{\lambda _{2}}_{i_{2}}\cdots x^{\lambda _{r}}_{i_{r}}\quad,\nonumber\\
t_{\alpha } & = & \prod _{k}\left( \sum _{j}x_{j}^{k}\right) ^{\alpha _{k}}\quad.
\end{eqnarray}
The numbers $\lambda $ denote a  partition, and the $\alpha $ label can be
related to a partition by
\begin{equation}
\lambda =(n^{\alpha _{n}}(n-1)^{\alpha _{n-1}}\cdots 2^{\alpha _{2}}1^{\alpha _{1}})\quad.
\end{equation}
For instance,  $\alpha =(1020\cdots 0)$ corresponds to $ \lambda =(331)$.

From the relations between the different representations one can
show that every excited state can be expressed as a product over 
traces.
An excited state $s_{\lambda }$ of the ${SU}(N)$ one-plaquette problem
is a partition of length less than $N$. If we use a lexicographical ordering
$\succ $ 
\cite{McD79},
it means that  $S_\lambda$
 contains only monomials
$m_{\kappa }$ where $\kappa \preceq \lambda $. It is easy to see
that a trace function $t_{\nu }$ only contains 
monomials $m_{\kappa }$
where $\kappa \succeq \nu $. Therefore, an excited state $\Psi _{\lambda }$
from Eq.~(\ref{E2}), can be expressed in terms of the 
$N-1$ independent trace variables, where 
\begin{equation}
t_{\alpha _{1}\alpha _{2}\cdots \alpha _{N-1}}=\trace [U]^{\alpha _{1}}\trace [U^{2}]^{\alpha _{2}}\cdots \trace [U^{N-1}]^{\alpha _{N-1}}\quad.
\end{equation}

We can thus express the anti-symmetric wave functions
in terms of the symmetric trace variables. 
One can easily construct the transition,
or Kostka,  tables \cite{McD79}
for transformations between $s$ and $m$, as well as $t$ and $m$.

Even though it is relatively straightforward to use trace variables
for $SU(2)$, it turns out to be rather involved to extend these
relations to higher-dimensional gauge theories. For example, with the
relations above we can determine the integration measure for trace
variables.  For $SU(2)$ we require the relations up to the partitions
of 4, while for $SU(3)$ we require similar relations for the
partitions of 6,
\begin{eqnarray}
SU(2):\quad\quad \Xi^2 & = & 1-t_{1}^{2}\quad,\\
SU(3):\quad\quad \Xi^2 & = & 27-5t_{1}^{3}+9t_{1}t_{2}-\frac{1}{2}t_{2}^{3}+\frac{5}{4}t_{1}^{2}t_{2}^{2}-t_{1}^{4}t_{2}+\frac{1}{4}t_{1}^{6}\\
 & = & 27-18t_{1}t_{-1}+4(t_{1}^{3}+t_{-1}^{3})-t_{1}^{2}t_{-1}^{2}\quad.
\end{eqnarray}
 where $t_{p}=\trace [U^{p}]$. However, the use for this relation for $SU(3)$
is restricted as the two complex trace variables $t_{1}$ and $t_{2}$
have complicated relations, and are restricted to a non-trivial domain of the
complex space ${\textbf {C}}^{2}$. Note that there are only two independent
angular variables for $SU(3)$, which makes the angular variables much more useful
than the trace variables.

The Haar measure $\mu (U)$ reduces to the measure of the trace variables
if the integrand depends only on the trace variables, 
\begin{equation}
\int d\mu (U)F(\{t_{\lambda }\})=\int \Xi dt_{1}\cdots dt_{N-1}F(\{t_{\lambda }\})=\int \Xi^2d\phi _{1}\cdots d\phi _{N-1}F(\{t_{\lambda }(\{\phi _{i}\})\})\quad,
\end{equation}
 where $\Xi$ is a non-polynomial function of the trace variables
$t_{i}$.

\subsection{The magnetic term}

The magnetic terms in coordinate representation are just the multiplication of
the state with $s_{1}$ and $s_{11\cdots 1}$, where $11\cdots 1$
is $1^{N-1}$. We can apply the Littlewood-Richardson rule for the general
multiplication of two Schur functions, 
\begin{equation}
s_{\lambda }s_{\kappa }=\sum _{\mu }c^{\mu }_{\lambda \kappa }s_{\mu }\quad.
\end{equation}
 However, since $s_{1}$ and $s_{11\cdots 1}$ are relatively simple we
 can derive multiplication rules from first principles, taking into
 account that the partitions cannot be longer than $N-1$. As symmetric
 polynomials we find ($x_{j}=\exp \{i\psi _{j}\}$),
\begin{eqnarray}
\trace [U] & = & x_{1}+x_{2}+x_{3}+\cdots +x_{N}\quad,\\
\trace [U^{-1}] & = & x_{1}x_{2}\cdots x_{N-1}+\cdots +x_{1}x_{3}\cdots x_{N}+x_{2}x_{3}\cdots x_{N}\quad,
\end{eqnarray}
 where we use that $x_{1}x_{2}\cdots x_{N}=1$. In terms of the partition
$\lambda $ of $s_{\lambda }$, multiplying with $\trace [U]$ means
adding one to each in turn $\lambda _{i}\rightarrow \lambda _{i}+1$, cancelling
all the partitions for which $\lambda _{i}>\lambda _{j}$ for $j<i$,
due to the intrinsic anti-symmetry. Multiplying with $\trace [U^{-1}]$
corresponds to $\lambda _{i}\rightarrow \lambda _{i}-1$, with the same
cancellations.

This relation can be generalised to the multiplication by $\trace
[U^{p}]$ for any power $p$. This reduces to adding $p$ to each
of the elements $\lambda _{i}$ in the partition $\lambda $. The
partitions should be ordered $\lambda _{1}>\lambda _{2}>\cdots
>\lambda _{N-1}>0$, and the non-distinct partitions dropped.

\section{Equivalent forms of the electric energy}

In this section we prove that the electric operator in angular
coordinates $\{\psi _{i}\}$,
\begin{equation}
D^s_N = -\frac{1}{2} \sum_{i=1}^N \frac{\partial^2}{\partial \psi_i^2}   +
\frac{1}{2} 
\left(\frac{1}{\sqrt{N}} \sum_{i=1}^N 
\frac{\partial}{\partial \psi_i} \right)^2 \quad ,
\end{equation}
the $SU(N)$ equivalent of Eq.~(\ref{Dn}), 
Can be realised in the trace representation $\{\trace [U^{p}]\}$
by the $SU(N)$ version of
the electric operator in Eq.~(\ref{E2}). 

We start of by the relation between a function $F$ of the trace variables and
one of the angular variables,
\begin{equation}
F\left( \trace [U],\frac{1}{2}\trace [U^{2}],\cdots ,\frac{1}{N-1}
\trace [U^{N-1}]\right) =F\left( \sum _{i=1}^{N}e^{i\psi _{i}},\frac{1}{2}
\sum _{i=1}^{N}e^{i2\psi _{i}},\cdots ,\frac{1}{N-1}
\sum _{i=1}^{N}e^{i(N-1)\psi _{i}}\right) \quad.
\end{equation}
In order to prove the fundamental relation, we need to know the
derivatives of the Vandermonde determinant with respect to the angular
variables
\begin{equation}
\Delta_{j}=\frac{1}{\Delta}\frac{\partial }{\partial \psi_{j}}\Delta =
\sum_{k\not =j}^{N}\frac{ie^{i\psi _{j}}}{e^{\psi _{j}}-e^{\psi _{k}}}\quad.
\end{equation}
 Writing out the terms in Eq.~(\ref{Dn}) we generate several terms, 
\begin{equation}
\frac{1}{\Delta}D^s_{N}\Delta F=\boxed {\frac{1}{\Delta}[D^s_{N}\Delta]F}_{{\, \, I}}+\boxed {D^s_{N}F}_{{\, \, II}}-\boxed {\sum _{i=1}^{N}\Delta_{i}\partial _{i}F}_{{\, \, III}}+\boxed {\frac{1}{N}\left( \sum _{i=1}^{N}\Delta_{i}\right) \left( \sum _{i=1}^{N}\partial _{i}F\right) }_{{\, \, IV}}\quad,
\end{equation}
 where $\partial _{i}=\frac{\partial }{\partial \psi _{i}}$. Some tedious,
but straightforward, calculations lead to 
\begin{eqnarray}
\mathrm{I} & = & \frac{N(N^{2}-1)}{24}F\quad,\\
\mathrm{II} & = & \frac{1}{2}\left( \sum _{p,q=1}^{N-1}\trace [U^{p+q}]F_{,pq}+\sum _{p=1}^{N-1}p\trace [U^{p}]F_{,p}\right) \\
 & - & \frac{1}{2N}\left( \sum _{p,q=1}^{N-1}\trace [U^{p}]\trace [U^{q}]F_{,pq}+\sum _{p=1}^{N-1}p\trace [U^{p}]F_{,p}\right) \quad,\\
\mathrm{III} & = & \frac{1}{2}\sum _{p=1}^{N-1}\left( \sum _{q=1}^{p-1}\frac{1}{2}\trace [U^{p-q}]\trace [U^{q}]+\left( N-\frac{1}{2}-\frac{p}{2}\right) \trace [U^{p}]\right) F_{,p}\quad,\\
\mathrm{IV} & = & -\frac{N-1}{2}\sum _{p=1}^{N-1}\trace [U^{p}]F_{,p}\quad,
\end{eqnarray}
 which, when substituted back in the original equation, yields the $SU(N)$
version of Eq.~(\ref{E2}):
\begin{eqnarray}
E^2_l F \left(\left\{ \frac{1}{p} {\rm Tr}[U^p] \right\}_{p}
 \right) & = &
\sum_{p,q}
\left( \frac{1}{2}{\rm Tr}[U^p U^q] -
\frac{1}{2N} {\rm Tr}[U^p] {\rm Tr}[U^q] \right) 
F_{,pq} \cr
& + & \sum_{p} \left( \frac{N^2-p}{2N} {\rm Tr}[U^p]
+ \frac{1}{2} \sum_{q=1}^{p-1} {\rm Tr}[U^{p-q}] {\rm Tr}[U^q]
\right) F_{,p} \ \ .
\end{eqnarray}
This finishes the proof.




\begin{thebibliography}{10}
\bibitem{YangMills} C. N. Yang and R. L. Mills,  \textit{Phys.~Rev.} {\bf 96} (1954), 191.
\bibitem{FS80}L. D. Faddeev and A. A. Slavnov, ``{Gauge Fields, Introduction to Quantum Theory},''
Benjamin Cummings, Reading (Mass.), 1980. 
\bibitem{Dir67}P. A. M.  Dirac, ``{Principles of Quantum Mechanics}, ''
4th edition, Oxford University Press, Oxford, 1967. 
\bibitem{Dirac}P. A. M. Dirac, ``{Lectures on Quantum Mechanics},'' Belfer Graduate School of Science (Yeshiva University), New York, 1964;
P. A. M. Dirac, ``{Lectures on Quantum Field Theory},'' 
Belfer Graduate School of Science (Yeshiva University), New York, 1966.
\bibitem{Mut87}T. Muta, ``{Foundations of Quantum Chromodynamics},'' World Scientific, Singapore, 1987. 
\bibitem{HT92}M. Henneaux and C. Teitelboim, ``{Quantization of Gauge Systems},'' Princeton University Press, Princeton, 1992. 
\bibitem{KS75}J. Kogut and L. Susskind, \textit{Phys.~Rev.~D} \textbf{11} (1975), 395. 
\bibitem{Kog79}J. B. Kogut, \textit{Rev.~Mod.~Phys.} \textbf{51} (1979), 659. 
\bibitem{Ber66}F. A. Berezin, ``{The Method of Second Quantization},'' Academic Press, New York, 1966. 
\bibitem{Fel81}B. Felsager, ``{Geometry, Particles and Fields},'' Odense University Press, Odense,
1981. 
\bibitem{'tHooft} G. 't Hooft and M. Veltman, \textit{Nucl.~Phys.~B} {\bf 44} (1972), 189. 
\bibitem{WilsonA}K. Wilson, \textit{Phys.~Rev.} {\bf 179} (1969), 499.
\bibitem{MM94}I. Montvay and G. M\"{u}nster, ``{Quantum Fields on a Lattice},'' Cambridge Univ.\ Press, Cambridge, 1994. 
\bibitem{Baa85}B. E. Baaquie, \textit{Phys.~Rev.~D} \textbf{32} (1985), 2774;
\textit{Phys.~Rev.~D} \textbf{33} (1986), 2367.
\bibitem{Sto94}D. Stoll, \textit{Phys.~Lett.~B} \textbf{ 336} (1994), 524. 
\bibitem{Hay98}R. W. Haymaker, preprint \textbf{hep-lat/9809094}. 
\bibitem{CL80}N. H. Christ and T. D. Lee, \textit{Phys.~Rev.~D} {\bf 22} (1980), 939.
\bibitem{MR84}V. F. M\"{u}ller and W. R\"{u}hl, \textit{Nucl.~Phys.~B} \textbf{230} (1984), 49. 
\bibitem{Bro85}J.~B. Bronzan, \textit{Phys.~Rev.~D} \textbf{31} (1985), 2020. 
\bibitem{Bro88}J. B. Bronzan, \textit{Phys.~Rev.~D} \textbf{37} (1988), 1621. 
\bibitem{Bro88b}J. B. Bronzan, \textit{Phys.~Rev.~D} \textbf{38} (1988), 1994. 
\bibitem{BV91}J. B. Bronzan and T. E. Vaughan, \textit{Phys.~Rev.~D} \textbf{43} (1991), 3499. 
\bibitem{BV93}J. B. Bronzan and T. E. Vaughan, \textit{Phys.~Rev.~D} \textbf{47} (1993), 3543. 
\bibitem{Man79}S. Mandelstam, \textit{Phys. Rev. D} \textbf{19} (1979), 2391. 
\bibitem{LenzNausThies}F. Lenz, H. W. L.  Naus, and M. Thies, \textit{Ann.~Phys.~(N.Y.)}
{\bf 233} (1994), 317.
\bibitem{Wat94}N. J. Watson, \textit{Phys.~Lett.~B} \textbf{323} (1994), 385. 
\bibitem{HK84}D. Horn and M. Karliner, \textit{Nucl.~Phys.~B} \textbf{235} (1984), 135. 
\bibitem{BR65}M. A. B. B\'{e}g and H. Ruegg, \textit{J.~Math.~Phys.}
\textbf{6} (1965), 677. 
\bibitem{Cre78}M. Creutz, \textit{J.~Math.~Phys.} \textbf{19} (1974), 2043. 
\bibitem{Cre83}M. Creutz, ``{Quarks, Gluons, and Lattices},''
 Cambridge Univ.\ Press, Cambridge, 1983. 
\bibitem{Leo98}C. R. Leonard, Ph.D. thesis, Melbourne University, 1998. 
\bibitem{Schuette}D. Sch\"{u}tte, Zheng Weihong, and C. J. Hamer,
\textit{Phys.~Rev.~D} {\bf 55} (1997), 2974.
\bibitem{bgl89}C. Di Bartolo, R. Gambini, and L. Leal, \textit{Phys.~Rev.~D} \textbf{39} (1989), 1756. 
\bibitem{MOS66}W. Magnus, F. Oberhettinger, and R. P. Soni, ``{Formulas and Theorems for
the Special Functions of Mathematical Physics},'' Springer, Berlin, 1966. 
\bibitem{CCMpapers} R. F. Bishop, \textit{Theor.~Chim.~Acta} {\bf 80} (1991), 95; \textit{in} ``{Microscopic Quantum Many-Body Theories and Their Application},''
 J. Navarro and A. Polls (eds.), Lec.~Notes in Physics \textbf{510}, Springer, Berlin, 1998, p. 1.
\bibitem{Het93}J. E. Hetrick, \textit{Int.~J.~Mod.~Phys.} \textbf{A9} (1994), 3153. 
\bibitem{Hal93}J. Hallin, \textit{Class.~Quant.~Grav.} \textbf{11} (1994), 1615. 
\bibitem{Weyl}H.~Weyl, ``{The Classical Groups},'' 2nd edition, Princeton University
Press, Princeton, 1946.
\bibitem{YMcirc}J. A. Minahan and A. P. Polychronakos, \textit{Phys.~Lett.~B} {\bf 312} (1993), 155;
 \textit{Phys.~Lett.~B} {\bf 326} (1994), 288; Interacting Fermion Systems
 from Two-Dimensional QCD, CERN preprint, 1993 ; M. R. Douglas, {\bf
 hep-th}/9311130; \textit{Nucl.~Phys.~Proc.~Suppl.} {\bf 41} (1995), 66;
 M. Caselle, A. D'Adda, L. Magnea, and S. Panzeri, \textit{Nucl.~Phys.~B} {\bf
 416} (1994), 751.
\bibitem{RingSchuck} P. Ring and P. Schuck, ``{The Nuclear
Many-Body Problem},'' Springer, New York, 1980.
\bibitem{Moshinski}M. Moshinski and Y. F. Smi, ``{The Harmonic Oscillator 
in Modern Physics },''
Harwood Academic, Amsterdam, 1996, (Contemporary concepts in physics, Vol.
{\bf 9}). 
\bibitem{RW80}D. Robson and D. M. Webber, \textit{Z.~Phys.~C} \textbf{7} (1980), 53. 
\bibitem{BlaizotRipka} J. P. Blaizot and G. Ripka, ``{Quantum Theory
of Finite Systems},'' MIT Press, Cambridge (Mass.), 1986.
\bibitem{WW27}E. T. Whittaker and G. N. Watson, ``{A Course in Modern Analysis},'' Cambridge Press, Cambridge, 1927. 
\bibitem{AS64}M. Abramowitz and I. A. Stegun, ``{Handbook of Mathematical Functions},'' National Bureau of Standards,
Dover, New York, 1965. 
\bibitem{Gre87}J. Greensite, \textit{Phys. Lett.  B} \textbf{191} (1987), 431. 
\bibitem{LWB}N. E. Ligterink, N. R. Walet, and R. F. Bishop, 
\textit{Ann. Phys. (N.Y.)} {\bf 267} (1998), 97.
\bibitem{CCMFT}  S. J. Baker, R. F. Bishop, and N. J. Davidson,
 \textit{Phys. Rev. D} {\bf 53} (1996), 2610;
S. J. Baker, R. F. Bishop, and N. J. Davidson, \textit{Nucl. Phys. B}
               (Proc. Supp.) {\bf 53} (1997), 834.
\bibitem{CCMmagnetism}R. F.  Bishop, J. B. Parkinson, and Yang Xian, \textit{Phys. Rev. B }
{\bf 44} (1991), 9425; D. J. J.  Farnell, R. F.  Bishop, and Chen, Zheng, \textit{J. Stat. Phys.}
{\bf 90} (1998) 327; J. Rosenfeld, N. E. Ligterink, and R. F. Bishop, \textit{Phys. Rev. B}
{\bf 60} (1999), 4030.
\bibitem{GL97}S.-H. Guo and X.-Q. Luo, preprint \textbf{hep-lat/9706017}. 
\bibitem{McD79}I. G. MacDonald, ``{Symmetric Polynomials and Hall Functions},'' Clarendon
Press, Oxford, 1979. 
\end{thebibliography}
\end{document}